\documentclass[sigconf]{acmart}
\usepackage{bm}
\usepackage{dsfont}
\usepackage{xspace}
\usepackage{grffile}
\usepackage{balance}

\def\BibTeX{{\rm B\kern-.05em{\sc i\kern-.025em b}\kern-.08emT\kern-.1667em\lower.7ex\hbox{E}\kern-.125emX}}

\copyrightyear{2019} 
\acmYear{2019} 
\setcopyright{acmlicensed}
\acmConference[SIGIR '19]{Proceedings of the 42nd International ACM SIGIR Conference on Research and Development in Information Retrieval}{July 21--25, 2019}{Paris, France}
\acmBooktitle{Proceedings of the 42nd International ACM SIGIR Conference on Research and Development in Information Retrieval (SIGIR '19), July 21--25, 2019, Paris, France}
\acmPrice{15.00}
\acmDOI{10.1145/3331184.3331259}
\acmISBN{978-1-4503-6172-9/19/07}

\begin{document}
\newcommand{\figscale}{.43}
\newcommand{\figscaletwo}{.42}
\newcommand\iverson[1]{\mathbb{I}\!\left[#1\right]}

\title[Statistical Significance Testing in Information Retrieval]{Statistical Significance Testing in Information Retrieval:\\An Empirical Analysis of Type I, Type II and Type III Errors}

\author{Juli\'an Urbano}
\affiliation{%
  \institution{Delft University of Technology}
  \city{The Netherlands}
}\email{urbano.julian@gmail.com}

\author{Harlley Lima}
\affiliation{%
  \institution{Delft University of Technology}
  \city{The Netherlands}
}\email{h.a.delima@tudelft.nl}

\author{Alan Hanjalic}
\affiliation{%
  \institution{Delft University of Technology}
  \city{The Netherlands}
}\email{a.hanjalic@tudelft.nl}

\begin{abstract}
Statistical significance testing is widely accepted as a means to assess how well a difference in effectiveness reflects an actual difference between systems, as opposed to random noise because of the selection of topics.
According to recent surveys on SIGIR, CIKM, ECIR and TOIS papers, the $t$-test is the most popular choice among IR researchers. However, previous work has suggested computer intensive tests like the bootstrap or the permutation test, based mainly on theoretical arguments. On empirical grounds, others have suggested non-parametric alternatives such as the Wilcoxon test.
Indeed, the question of which tests we should use has accompanied IR and related fields for decades now. Previous theoretical studies on this matter were limited in that we know that test assumptions are not met in IR experiments, and empirical studies were limited in that we do not have the necessary control over the null hypotheses to compute actual Type I and Type II error rates under realistic conditions. Therefore, not only is it unclear which test to use, but also how much trust we should put in them.
In contrast to past studies, in this paper we employ a recent simulation methodology from TREC data to go around these limitations. Our study comprises over 500 million $p$-values computed for a range of tests, systems, effectiveness measures, topic set sizes and effect sizes, and for both the 2-tail and 1-tail cases. Having such a large supply of IR evaluation data with full knowledge of the null hypotheses, we are finally in a position to evaluate how well statistical significance tests really behave with IR data, and make sound recommendations for practitioners.
\end{abstract}


\maketitle

\sloppy
\section{Introduction}\label{sec:intro}

In the traditional test collection based evaluation of Information Retrieval (IR) systems, statistical significance tests are the most popular tool to assess how much noise there is in a set of evaluation results. Random noise in our experiments comes from sampling various sources like document sets \citep{robertson2012vvariance,sanderson2012differences,jones2014size} or assessors \citep{voorhees2000variations,bailey2008judges,alonso2012crowdsourcing}, but mainly because of topics \citep{voorhees2002size,carterette2008thousands,urbano2013measurement,sakai2016design,urbano2016reliability}. Given two systems evaluated on the same collection, the question that naturally arises is ``\textsl{how well does the observed difference reflect the real difference between the systems and not just noise due to sampling of topics}''? Our field can only advance if the published retrieval methods truly outperform current technology on unseen topics, as opposed to just the few dozen topics in a test collection. 
Therefore, statistical significance testing plays an important role in our everyday research to help us achieve this goal.

\subsection{Motivation}

In a recent survey of 1,055 SIGIR and TOIS papers, \citet{sakai2016statistical} reported that significance testing is increasingly popular in our field, with about 75\% of papers currently following one way of testing or another. In a similar study of 5,792 SIGIR, CIKM and ECIR papers, \citet{carterette2017but} also observed an increasing trend with about 60\% of full papers and 40\% of short papers using significance testing. The most typical situation is that in which two IR systems are compared on the same collection, for which a paired test is in order to control for topic difficulty. According to their results, the $t$-test is the most popular with about 65\% of the papers, followed by the Wilcoxon test in about 25\% of them, and to a lesser extent by the sign, bootstrap-based and permutation tests.

It appears as if our community has made a de facto choice for the $t$ and Wilcoxon tests. Notwithstanding, the issue of statistical testing in IR has been extensively debated in the past, with roughly three main periods reflecting our understanding and practice at the time. In the 1990s and before, significance testing was not very popular in our field, and the discussion largely involved theoretical considerations of classical parametric and non-parametric tests, such as the $t$-test, Wilcoxon test and sign test \citep{vanrijsbergen1979ir,hull1993statistical}.
During the late 1990s and the 2000s, empirical studies became to be published, and suggestions were made to move towards resampling tests based on the bootstrap or permutation tests, while at the same time advocating for the simplicity of the $t$-test \citep{savoy1997statistical,zobel1998how,sanderson2005effort,smucker2007comparison}.
Lastly, the 2010s witnessed the wide adoption of statistical testing by our community, while at the same time it embarked in a long-pending discussion about statistical practice at large \cite{carterette2012mcp,carterette2015bayesian,sakai2014reform}.

Even though it is clear that statistical testing is common nowadays, the literature is still rather inconclusive as to which tests are more appropriate.
Previous work followed empirical and theoretical arguments. Studies of the first type usually employ a topic splitting methodology with past TREC data. The idea is to repeatedly split the topic set in two, analyze both separately, and then check whether the tests are concordant (i.e. they lead to the same conclusion). These studies are limited because of the amount of existing data and because typical collections only have 50 topics, so we can only do 25--25 splits, abusing the same 50 topics over and over again. But most importantly, without knowledge of the null hypothesis we can not simply count discordances as Type I errors; the tests may be consistently wrong or right between splits.
On the other hand, studies of theoretical nature have argued that IR evaluation data does not meet test assumptions such as normality or symmetry of distributions, or measurements on an interval scale~\citep{ferrante2017ir}. In this line, it has been argued that the permutation test may be used as a reference to evaluate other tests, but again, a test may differ from the permutation test on a case by case basis and still be valid in the long run with respect to Type I errors\footnote{For instance, under $H_0$ a test that always returns 1 minus the $p$-value of the permutation test, will always be discordant with it but have the same Type I error rate.}. 

\subsection{Contributions and Recommendations}

The question we ask in this paper is: \textsl{which is the test that, maintaining Type I errors at the $\alpha$ level, achieves the highest statistical power with IR-like data?} In contrast to previous work, we follow a simulation approach that allows us to evaluate tests with unlimited IR-like data and under full control of the null hypothesis. In particular, we use a recent method for stochastic simulation of evaluation data~\citep{urbano2018stochastic,urbano2016reliability}. In essence, the idea is to build a generative stochastic model of the joint distribution of effectiveness scores for a pair of systems, so that we can simulate an arbitrary number of new random topics from it. The model contains, among other things, the true distributions of both systems, so we have full knowledge of the true mean scores needed to compute test error rates.

We initially devised this simulation method to study the problem of topic set size design~\citep{urbano2016reliability}, but after further developments we recently presented preliminary results about the $t$-test with IR data~\citep{urbano2018stochastic}. The present paper provides a thorough study in continuation. Our main \textbf{contributions} are as follows:
\begin{itemize}
\item A description of a methodology that allows us, for the first time, to study the behavior of statistical tests with IR data.
\item A large empirical study of 500 million $p$-values computed for simulated data resembling TREC Adhoc and Web runs.
\item A comparison of the typical paired tests ($t$-test, Wilcoxon, sign, bootstrap-shift and permutation) in terms of actual Type I, Type II and Type III errors\footnote{Type III errors refer to incorrect directional conclusions due to correctly rejected non-directional hypotheses (ie. correctly rejecting $H_0$ for the wrong reason)~\citep{kaiser1960directional}.}.
\item A comparison across several measures ($AP$, $nDCG@20$, $ERR@20$, $P@10$ and $RR$), topic set sizes (25, 50 and 100), significance levels (0.001--0.1), effect sizes (0.01--0.1), and for both the 2-tailed and the 1-tailed cases.
\end{itemize}

Based on the results of this study, we make the following conclusions and \textbf{recommendations}:
\begin{itemize}
\item The $t$-test and the permutation test behave as expected and maintain the Type I error rate at the $\alpha$ level across measures, topic set sizes and significance levels. Therefore, our field is \emph{not} being conservative in terms of decisions made on the grounds of statistical testing.
\item Both the $t$-test and the permutation test behave almost identically. However, the $t$-test is simpler and remarkably robust to sample size, so it becomes our top recommendation. The permutation test is still useful though, because it can accommodate other test statistics besides the mean.
\item The bootstrap-shift test shows a systematic bias towards small $p$-values and is therefore more prone to Type I errors. Even though large sample sizes tend to correct this effect, we propose its discontinuation as well.
\item We agree with previous work in that the Wilcoxon and sign tests should be discontinued for being unreliable.
\item The rate of Type III errors is not negligible, and for measures like $P@10$ and $RR$, or small topic sets, it can be as high as 2\%. Testing in these cases should be carried with caution.
\end{itemize}

We provide online fast implementations of these tests, as well as all the code and data to replicate the results found in the paper\footnote{\texttt{https://github.com/julian-urbano/sigir2019-statistical}}.

\section{Related Work}\label{sec:related}

One of the first accounts of statistical significance testing in IR was given by \citet{vanrijsbergen1979ir}, with a detailed description of the $t$, Wilcoxon and sign tests. Because IR evaluation data violates the assumptions of the first two, the sign test was suggested as the test of choice. \citet{hull1993statistical} later described a wider range of tests for IR, and argued that the $t$-test is robust to violations of the normality assumption, specially with large samples. However, they did not provide empirical results. Shortly after, \citet{savoy1997statistical} stressed van Rijsbergen's concerns and proposed bootstrapping as a general method for testing in IR because it is free of distributional assumptions. 

An early empirical comparison of tests was carried out by~\citet{wilbur1994nonparametric}, who simulated system runs using a simple model of the rate at which relevant documents are retrieved~\citep{mccarn1990mathematical}. Albeit unrealistic, this model allowed a preliminary study of statistical tests under knowledge of the null hypothesis. They showed that non-parametric tests, as well as those based on resampling, behaved quite well in terms of Type I error rate, also indicating preference for bootstrapping. 
\citet{zobel1998how} compared the $t$-test, Wilcoxon and ANOVA with a random 25--25 split of TREC Ad hoc topics, and found lower discordance rates with the $t$-test. However, they suggested the Wilcoxon test because it showed higher power and it has more relaxed assumptions.
Although they did not study significance tests,~\citet{voorhees2002size} used TREC Ad hoc and Web data to analyze the rate of discordance given the observed difference between two systems.
Inspired by these two studies, \citet{sanderson2005effort} used several 25--25 splits of TREC Ad hoc topics and found that the $t$-test has lower discordance rates than the Wilcoxon test, followed by the sign test.
\citet{cormack2007ttest} later used 124--124 topic splits from the TREC Robust track to compare actual and expected discordance rates, and found the Wilcoxon test to be more powerful than the $t$ and sign tests, though with higher discordance rates.
\citet{voorhees2009redux} similarly used 50--50 splits of the TREC Robust topics to study concordance rates of the $t$-test with score standardization. Finally, \citet{sakai2006bootstrap} also advocated for the bootstrap when evaluating effectiveness measures, but did not compare to other tests.

The most comprehensive comparison of statistical tests for IR is probably the one by \citet{smucker2007comparison}. From both theoretical and empirical angles, they compared the $t$, Wilcoxon, sign, bootstrap-shift and permutation tests. From a theoretical standpoint, they recommend the permutation test and compare the others to it using TREC Ad hoc data. They propose the discontinuation of the Wilcoxon and sign tests because they tend to disagree, while the bootstrap and $t$-test show very high agreement with the permutation test. In a later study, they showed that this agreement is reduced with smaller samples, and that the bootstrap test appears to have a bias towards small $p$-values~\citep{smucker2009agreement}.
Inspired by \citet{voorhees2009redux} and \citet{smucker2007comparison}, \citet{urbano2013comparison} performed a large-scale study with 50--50 splits of the TREC Robust data to compare statistical tests under different optimality criteria. They found the bootstrap test to be the most powerful, the $t$-test the one with lowest discordance rates, and the Wilcoxon test the most stable.

In summary, we find in the literature:
\begin{itemize}
\item Both theoretical and empirical arguments for and against specific tests.
\item Even though discordance rates can not be used as proxies to the Type I error rate because the null hypothesis is unknown~\citep{cormack2007ttest}, several studies made a direct or indirect comparison, suggesting that tests are too conservative because discordance rates are below the significance level $\alpha=0.05$.
\item Most studies analyze statistical tests with $AP$ scores, with few exceptions also using $P@10$. In general, they found higher discordance rates with $P@10$, arguing that $AP$ is a more stable measure and conclusions based on it are more reliable.
\item Studies of 2-tailed tests at $\alpha=0.05$ almost exclusively.
\item Except \citep{sakai2016two} and \citep{wilbur1994nonparametric}, no empirical study was carried out with control of the null hypothesis. However, the first one does not study the paired test case, and the second was based on unrealistic generative models for simulation.
\end{itemize}

Although previous work substantially contributed to our understanding of significance testing in IR, a comprehensive study of actual error rates is still missing in our field. 
An attempt at filling this gap was very recently presented by~\citet{parapar2019using}. Their approach exploits score distribution models from which new relevance profiles may be simulated (ie.~ranked lists of relevance scores) with indirect control over the $AP$ score they are expected to produce. 
Therefore, their method simulates \textsl{new runs} for the \textsl{same topic}. However, the question of statistical testing in IR involves hypotheses about the performance of two systems on a population of topics from which the test collection is sampled. To study these hypotheses, one therefore needs to simulate \textsl{new topics} for the \textsl{same systems}. 
In contrast to~\citep{parapar2019using}, our method does simulate the effectiveness scores of the given systems on random topics, directly and with full control over the distributions. Based on this simulation method, we aim at making informed recommendations about the practice of statistical testing in IR.


\section{Methods}\label{sec:methods}

This section provides a brief description of the five statistical tests we consider, and then outlines the method we follow to simulate evaluation data with which to compute actual error rates.

\subsection{Statistical Tests}\label{sec:methods:tests}

Assuming some effectiveness measure, let $B_1,\dots,B_n$ be the scores achieved by a baseline system on each of the $n$ topics in a collection, and let $E_1,\dots,E_n$ be the scores achieved by an experimental system on the same topics. For simplicity, let $D_i=E_i-B_i$ be the difference for a topic, and let $\overline{B}, \overline{E}$ and $\overline{D}$ be the mean scores over topics.
Under normal conditions, researchers compute the mean scores, and if $\overline{D}>0$ they test for the statistical significance of the result using a paired test. At this point, a distinction is made between directional and non-directional null hypotheses.

A non-directional null hypothesis has the form $H_0:\mu_B\!=\!\mu_E$, meaning that the mean score of both systems is the same; the alternative is $H_1:\mu_B\!\neq\!\mu_E$. This is called non-directional because it tests for equality of means, but if the hypothesis is rejected it does not say anything about the direction of the difference. In this case, one uses a test to compute a 2-tailed or 2-sided $p$-value.
A directional null hypothesis has the form $H_0:\mu_B\!\geq\!\mu_E$, meaning that the mean performance of the baseline system is larger than or equal to that of the experimental system\footnote{$H_0:\mu_B\leq\mu_E$ is also valid, but irrelevant to an IR researcher.}; the alternative is $H_1:\mu_B\!<\!\mu_E$. In this case, one computes a 1-tailed or 1-sided $p$-value.

For simplicity, let us assume $\overline{D}>0$ unless otherwise stated.

\subsubsection{Student's t-test}

The case of a paired two-sample test for $\mu_B=\mu_E$ is equivalent to the one-sample test for $\mu_D=0$.
In general, the distribution of a sample mean has variance $\sigma^2/n$. When this mean is normally distributed, and $\sigma$ is unknown but estimated with the sample standard deviation $s$, the standardized mean follows a $t$-distribution with $n-1$ degrees of freedom~\citep{student1908probable}. The test statistic is therefore defined as
\begin{equation}
t=\frac{\overline{D}}{s_D / \sqrt{n}}. \label{eq:t}
\end{equation}
Using the \textit{cdf} of the $t$-distribution, the 1-tailed $p$-value is calculated as the probability of observing an even larger $t$ statistic:
\begin{equation}
p_1=1-F_t(t~;~n-1) .
\end{equation}
Because the $t$-distribution is symmetric, the 2-tailed $p$-value is simply twice as large: $p_2=2\cdot p_1$. When the data are normally distributed, one can safely use the $t$-test because the mean is then normally distributed too. If not, and by virtue of the Central Limit Theorem, it can also be used if the sample size is not too small. 

In our experiments we use the standard R implementation in function \texttt{t.test}.

\subsubsection{Wilcoxon Signed Rank test}

This is a non-parametric test that disregards the raw magnitudes of the differences, using their ranks instead~\citep{wilcoxon1945individual}. First, every zero-valued observation is discarded\footnote{Other ways of dealing with zeros have been proposed, but they are well out of the scope of this paper. See for instance~\citep{pratt1959remarks}, and \citep{conover1973methods} for a comparison.}, and every other $D_i$ is converted to its rank $R_i$ based on the absolute values, but maintaining the correct sign: $\text{sign}(R_i)=\text{sign}(D_i)$. The test statistic is calculated as the sum of positive ranks
\begin{equation}
W=\sum_{R_i>0}{R_i} .
\end{equation}
Under the null hypothesis, $W$ follows a Wilcoxon Signed Rank distribution with sample size $n_0$ (the number of non-zero observations). The 1-tailed $p$-value is computed as
\begin{equation}
p_1=1-F_\text{WSR}(W; n_0),
\end{equation}
and the 2-tailed $p$-value is simply twice as large: $p_2=2\cdot p_1$.

In our experiments we use the standard R implementation in function \texttt{wilcox.test}.

\subsubsection{Sign test}

In this case the data are looked at as if they were coin tosses where the possible outcomes are $D_i>0$ or $D_i<0$, therefore having complete disregard for magnitudes~\citep[\S3.4]{conover1999practical}. The test statistic is the number of successes, that is, the number of topics where $D_i>0$:
\begin{equation}
S=\sum{\iverson{D_i>0}},
\end{equation}
where $\iverson{\bullet}$ is 1 if $\bullet$ is true or 0 otherwise. Under the null hypothesis, $S$ follows a Binomial distribution with 50\% probability of success and $n_0$ trials, where $n_0$ is again the number of topics where $D_i\neq 0$. The 1-tailed $p$-value is the probability of obtaining at least $S$ successes:
\begin{equation}
p_1=1-F_\textit{Binom}(S-1; n_0, 0.5) .
\end{equation}
The 2-tailed $p$-value is simply twice as large: $p_2=2\cdot p_1$. 
\citet{vanrijsbergen1979ir} proposed to use a small threshold $h$ such that if $|D_i|\leq h$ then we consider that systems are tied for topic $i$. Following \citet{smucker2009agreement}, we set $h=0.01$.

In our experiments we use our own implementation to compute $S$ and $n_0$, and then use the standard R implementation in function \texttt{binom.test} to compute the $p$-values.

\subsubsection{Permutation test}

This test is based on the exchangeability principle: under the null hypothesis both systems are the same, so the two scores observed for a topic actually come from the same system and we simply happened to label them differently~\citep[\S II]{fisher1935design,pitman1937significance}. With $n$ topics, there are $2^n$ different permutations of the labels, all equally likely. The one we actually observed is just one of them, so the goal is to calculate how extreme the observed $\overline{D}$ is.

In practice, the distribution under the null hypothesis is estimated via Monte Carlo methods. In particular, the following may be repeated $T$ times. A permutation replica $D^*_j$ is created by randomly swapping the sign of each $D_i$ with probability 0.5 (i.e. permuting the labels), and the mean $\overline{D}^*_j$ is recorded. The 1-tailed $p$-value is computed as the fraction of replicas where the mean is as large as the observed mean:
\begin{equation}
p_1=\frac{1}{T}\sum_j{\iverson{ \overline{D}^*_j \geq \overline{D} }}. \label{eq:p:p1}
\end{equation}
The 2-tailed $p$-value is similarly the fraction of replicas where the magnitude of the mean is at least as large as the one observed:
\begin{equation}
p_2=\frac{1}{T}\sum_j{\iverson{ |\overline{D}^*_j| \geq |\overline{D}| }}. \label{eq:p:p2}
\end{equation}

The precision of the $p$-values depends on how many replicas we create. As noted by \citet[ch. 15]{efron1998bootstrap}, an answer to this issue may be given by realizing that $T\cdot p_1$ has a Binomial distribution with $T$ trials and probability of success $p_1$. The coefficient of variation for the estimated $\hat{p}_1$ is
\begin{equation}
cv(\hat{p}_1)=\sqrt{\frac{\hat{p}_1(1-\hat{p}_1)}{T}}.
\end{equation}
If we do not want our estimate of $p_1$ to vary more than $\varepsilon\%$ of its value, then we need to set $T=\frac{(1-p_1)}{\varepsilon^2p_1}$. For instance, for a target $\varepsilon=1\%$ error on a $p_1=0.05$ (i.e., an error of $0.0005$), we need $T=190,\!000$ replicas. Under symmetricity assumptions, for $p_2$ we may assume $p_2=2\cdot p_1$. \citet{smucker2009agreement} and \citet{parapar2019using} used 100,000 replicas in their experiments, which yield an error of 0.00045 for a target $p_2=0.01$ (4.5\%), and an error of 0.001 for a target $p_2=0.05$ (2\%). Because we want to study small significance levels, in our experiments we use $T=1$ million replicas. This yields an error of 0.00014 for a target $p_2=0.01$ (1.4\%), and 0.0003 for $p_2=0.05$ (0.62\%). For the 1-tailed case, errors are 0.0001 (1\%) and 0.0002 (0.43\%), respectively.

Because this test is computationally expensive, we used our own implementation in C++, using the modern Mersenne Twister pseudo-random number generator in the C++11 standard.

\subsubsection{Bootstrap test -- Shift method}

This test follows the bootstrap principle to estimate the sampling distribution of the mean under the null hypothesis~\citep[\S16.4]{efron1998bootstrap}. The following is repeated $T$ times. A bootstrap sample $D^*_j$ is created by resampling $n$ observations \emph{with} replacement from $\{D_i\}_{i=1}^n$, and its mean $\overline{D}^*_j$ is recorded. Let $\overline{D}^*=1/T\sum_j{\overline{D}^*_j}$ be the mean of these means, which will be used to shift the bootstrap distribution to have zero mean. The 1-tailed $p$-value is computed as the fraction of bootstrap samples where the shifted bootstrap mean is at least as large as the observed mean:
\begin{equation}
p_1=\frac{1}{T}\sum_j{\iverson{ \overline{D}^*_j-\overline{D}^* \geq \overline{D} }}. \label{eq:b:p1}
\end{equation}
The 2-tailed $p$-value is similarly the fraction of shifted bootstrap samples where the magnitude of the mean is at least as large as the one observed:
\begin{equation}
p_2=\frac{1}{T}\sum_j{\iverson{ |\overline{D}^*_j-\overline{D}^*| \geq |\overline{D}| }}. \label{eq:b:p2}
\end{equation}

As with the permutation test, we compute $T=1$ million bootstrap samples. \citet{wilbur1994nonparametric} and \citet{sakai2006bootstrap} used 1,000 samples, \citet{cormack2006statistical} used 2,000, \citet{savoy1997statistical} used up to 5,000, and both \citet{smucker2007comparison} and \citet{parapar2019using} used 100,000. Again, we use our own implementation in C++11 for efficiency.

\subsection{Stochastic Simulation}\label{sec:methods:simulation}

\newcommand{\sysB}{\textsf{B}\xspace}
\newcommand{\sysE}{\textsf{E}\xspace}

\begin{figure}[t]
\centering\includegraphics[scale=.46]{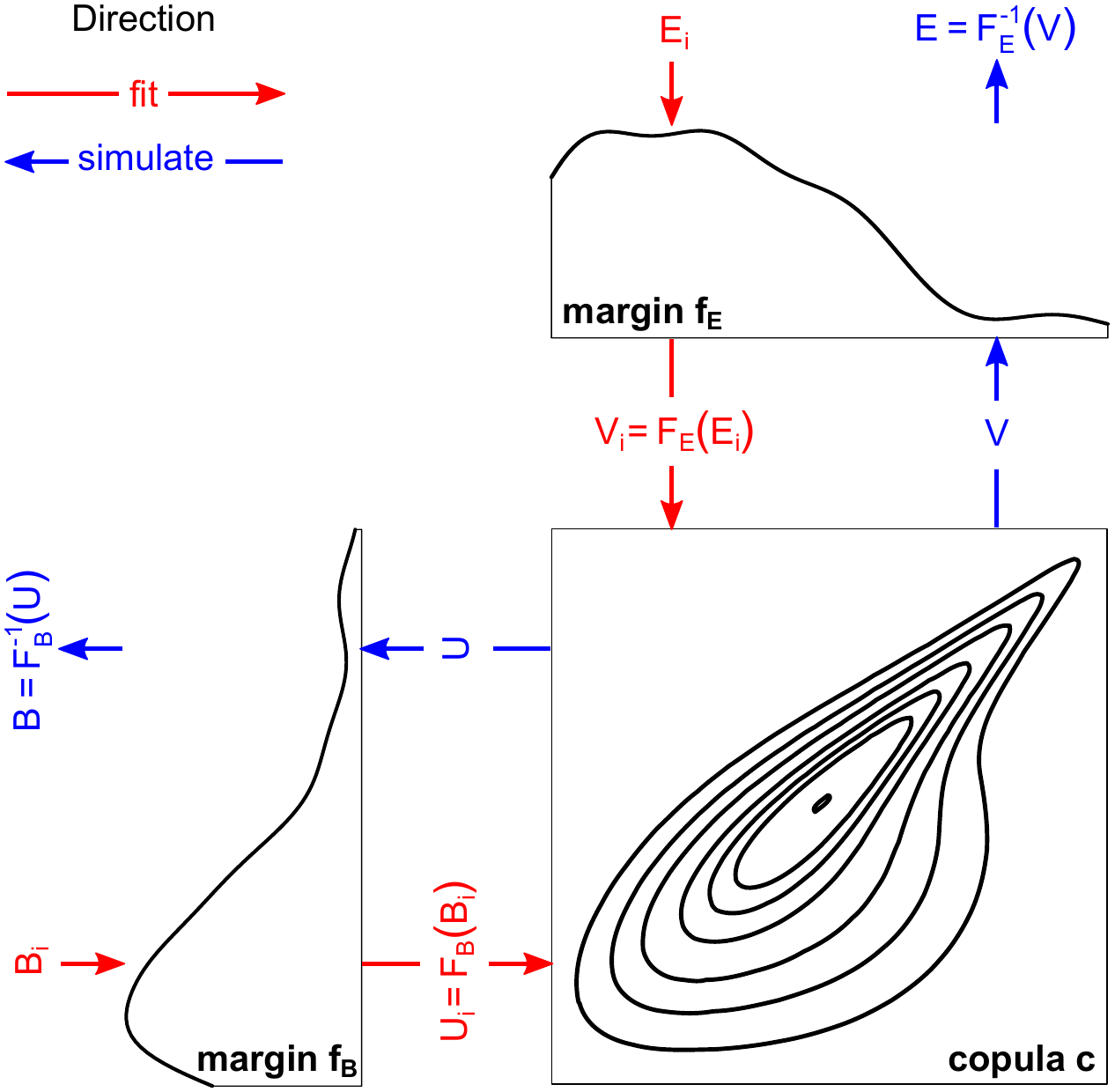}
\caption{Stochastic simulation model. To fit the model (red): 1) the per-topic scores $\bm{B_i}$ and $\bm{E_i}$ are used to fit the margin distributions $\bm{f_B}$ and $\bm{f_E}$, 2) the pseudo-observations $\bm{U_i}$ and $\bm{V_i}$ are computed with the \emph{cdf} functions, and 3) they are used to fit the copula $\bm{c}$. To simulate from the model (blue): 1) new pseudo-observations $\bm{U}$ and $\bm{V}$ are generated from the copula, and 2) the final scores $\bm{B}$ and $\bm{E}$ are computed with the inverse \emph{cdf} functions.}
\label{fig:flow}
\end{figure}

In order to simulate realistic IR evaluation that allows us to compare statistical tests, we use the method recently proposed by~\citet{urbano2018stochastic}, which extends an initial method by~\citet{urbano2016reliability} to study topic set size design methods. In essence, they build a generative stochastic model $M$ that captures the joint distribution of effectiveness scores for a number of systems. An observation from this model would be the vector of scores achieved by the system on the same topic. This model contains two parts: the marginal distribution of each system, that is, its distribution of effectiveness scores regardless of the other systems, and a copula~\citep{joe2014copulas} to model the dependence among systems, that is, how they tend to behave for the same topic. Because we will focus on paired statistical testing, in our case we will only contemplate bivariate models of only two systems \sysB and \sysE. 

Given a model $M$, we can indefinitely simulate evaluation scores for the same two systems but on new, random topics. The model defines the two individual marginal distributions $F_B$ and $F_E$, so it provides full knowledge about the null hypothesis because we know $\mu_B$ and $\mu_E$ beforehand. For the simulated data to be realistic, \citet{urbano2018stochastic} argue that the model first needs to be flexible enough to accommodate arbitrarily complex data; this is ensured by allowing a range of non-parametric models for both the margins and the copula. Finally, a realistic model can be instantiated by fitting it to existing TREC data using a model selection criterion that rewards fit over simplicity (e.g. log-likelihood instead of BIC). As they note, the resulting model is \emph{not} a model of the true performance of the existing systems whose data was used in the fitting process, but rather of \emph{some} hypothetical systems that behave similarly.

Figure~\ref{fig:flow} summarizes how these stochastic models are built and used. To fit the model (red flow), we start from the existing effectiveness scores of systems \sysB and \sysE over $n$ topics: the paired $\{(B_i,E_i)\}_{i=1}^n$ observations. The individual $B_1,\dots,B_n$ scores from the first system are used to fit the marginal distribution $F_B$, which determines the true mean score $\mu_B$, and likewise with the second system. The \textit{cdf} of the fitted margins are used to transform the original scores to pseudo-observations $\left\{\left(U_i=F_B(B_i), V_i=F_E(E_i)\right)\right\}_{i=1}^n$, such that $U_i,V_i\sim \textit{Uniform}$. The copula $c$ is then fitted to these pseudo-observations.
To simulate a new topic from the model (blue flow), a new pseudo-observation $(U,V)$ is randomly generated from the copula, which is then transformed back with the inverse \textit{cdf} to obtain the final observation $\left(B=F_B^{-1}(U), E=F_E^{-1}(V)\right)$. By construction, we have $B\sim F_B$ and $E\sim F_E$, so that we simulate new scores under full knowledge of the true system distributions. 

We use this simulation method in two different ways. First, in order to study Type I error rates we need to generate data under the null hypothesis $H_0:\mu_B\!=\!\mu_E$. We achieve this by assigning $F_E \leftarrow F_B$ after fitting the margins and the copula. This way, we simulate data from two systems that have a certain dependence structure but the same margin distribution, so that $\mu_B=\mu_E$ by construction.
Second, and in order to study Type II and Type III error rates, we need to generate data with different effect sizes. In particular, for a fixed difference $\delta$, we will need to make sure that $\mu_E=\mu_B +\delta$. Given an already fitted stochastic model, \citet[\S3.4]{urbano2018stochastic} show how this requirement can be met by performing a slight transformation of $F_E$ that still maintains underlying properties of the distribution such as its support. For instance, if $F_E$ corresponds to $P@10$ scores, the transformed distribution will also generate valid $P@10$ scores. This way, we simulate data from two systems that have a certain dependence structure and whose expected scores are at a fixed distance $\delta$. 

In this paper we use version 1.0 of the \texttt{simIReff} R package\footnote{\texttt{https://cran.r-project.org/package=simIReff}} by~\citet{urbano2018stochastic}. \texttt{simIReff} offers a high degree of flexibility to model IR data. In particular, it implements 6 distribution families for the margins (Truncated Normal, Beta, Truncated Normal Kernel Smoothing, Beta Kernel Smoothing, Beta-Binomial and Discrete Kernel Smoothing), and 12 copula families for the dependence structure (Gaussian, $t$, Clayton, Gumbel, Frank, Joe, BB1, BB6, BB7, BB8, Tawn 1 and Tawn 2) plus rotations. When transforming a distribution so that it has a certain expected value, we require a maximum absolute tolerance of $10^{-5}$.

\section{Evaluation}\label{sec:eval}


Because our goal is to evaluate the behavior of statistical tests on a variety of IR data, we chose systems and measures representative of ad hoc retrieval. In particular, we used the runs from the TREC 5--8 Ad hoc and TREC 2010--13 Web tracks. In terms of measures, we chose $AP$, $P@10$ and $RR$ for the Ad hoc runs, and $nDCG@20$ and $ERR@20$ for the Web runs.
We therefore evaluate tests on measures producing smooth but asymmetric distributions ($AP$, $nDCG@20$ and $ERR@20$), discrete ($P@10$), and with non-standard, non-uniform support ($RR$).

As is customary in this kind of experiments, we only use the top 90\% of the runs to avoid errorful system implementations. This results in a total of 326 Ad hoc runs and 193 Web runs, from which we can use about $14,\!000$ and $5,\!000$ run pairs, respectively. In terms of topic set sizes, we will simulate results on $n=25, 50, 100$ topics, representing small, typical and large collections.

\subsection{Type I Errors}\label{sec:results:type1}

\begin{figure*}[!ht]
\includegraphics[scale=\figscale]{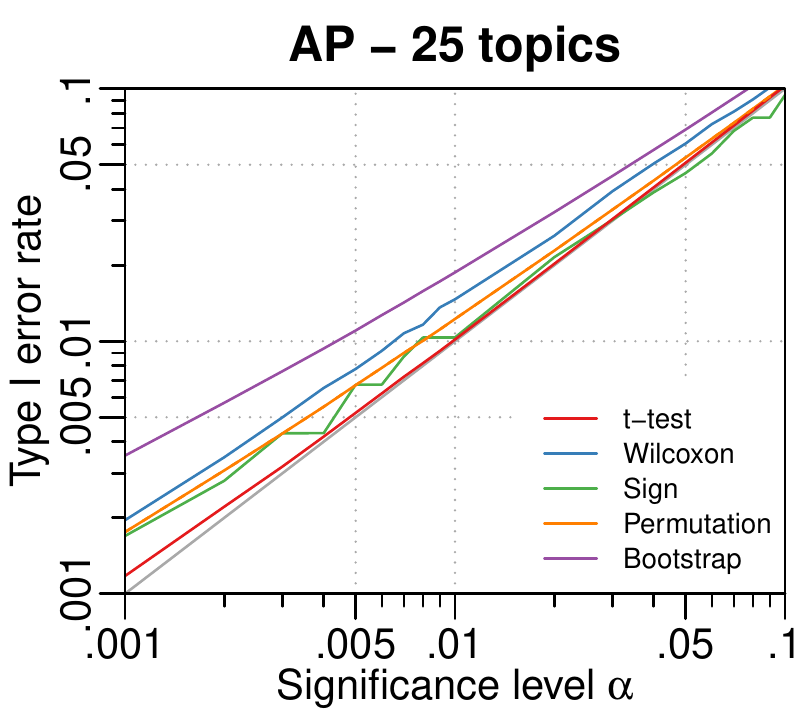}~\includegraphics[scale=\figscale]{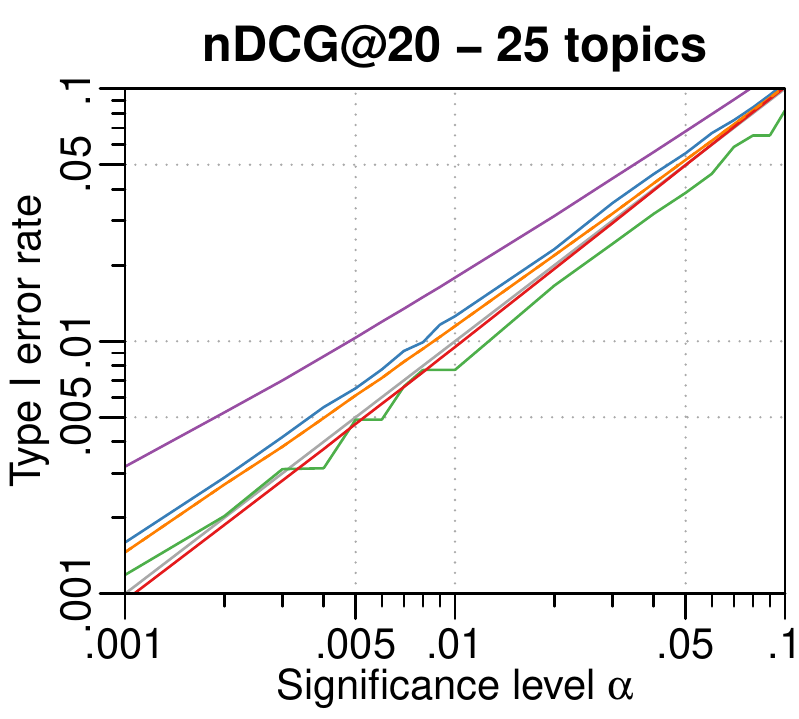}~\includegraphics[scale=\figscale]{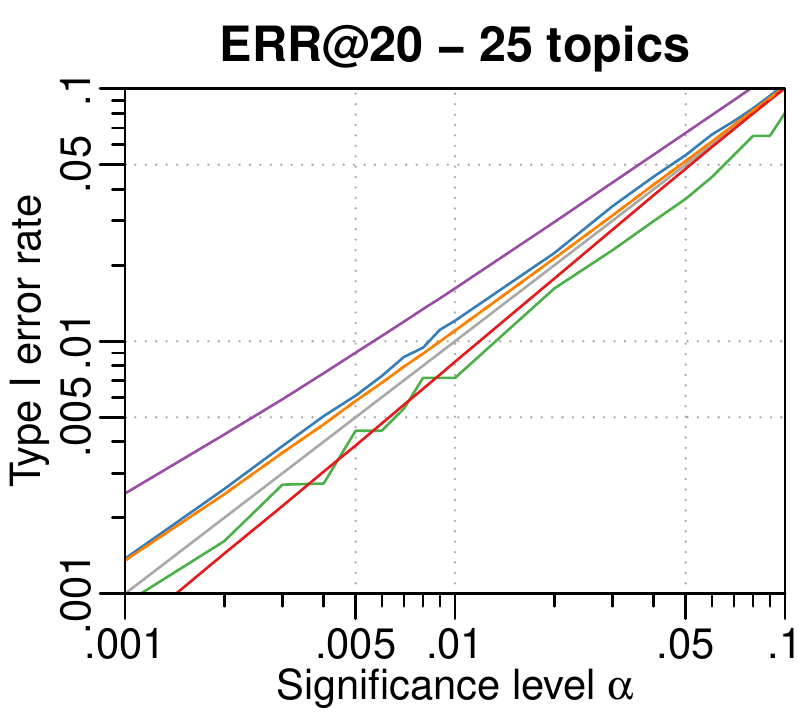}~\includegraphics[scale=\figscale]{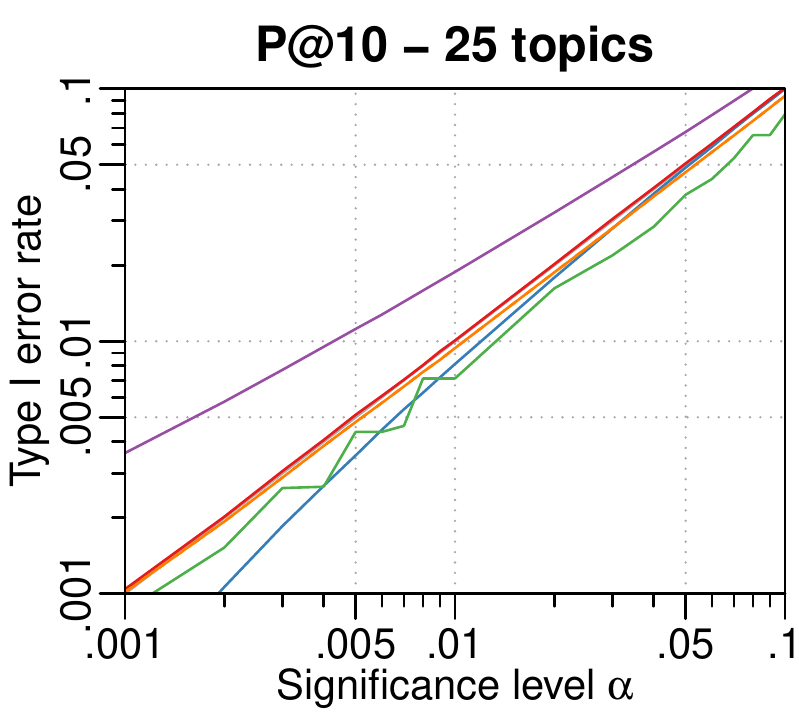}~\includegraphics[scale=\figscale]{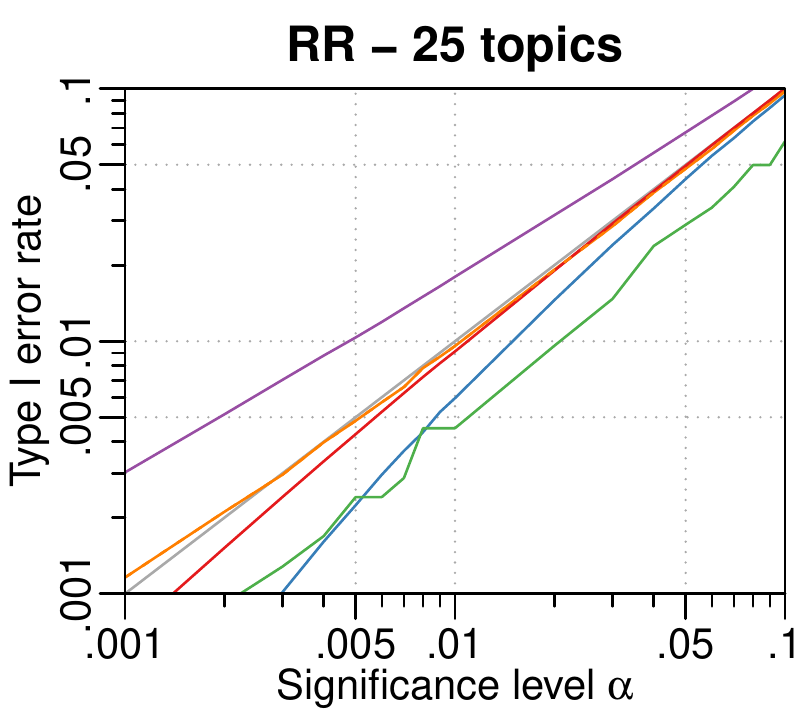}\\
\includegraphics[scale=\figscale]{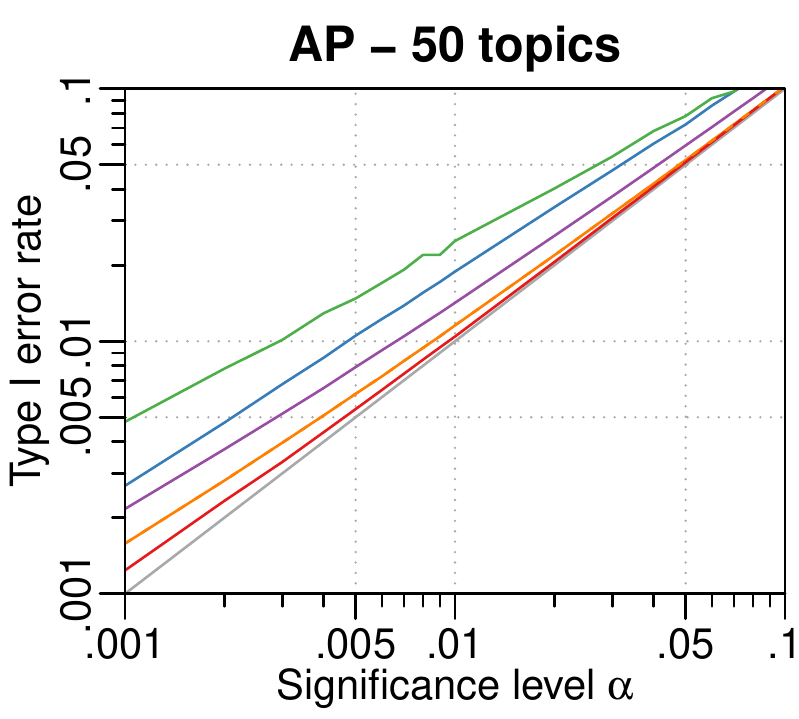}~\includegraphics[scale=\figscale]{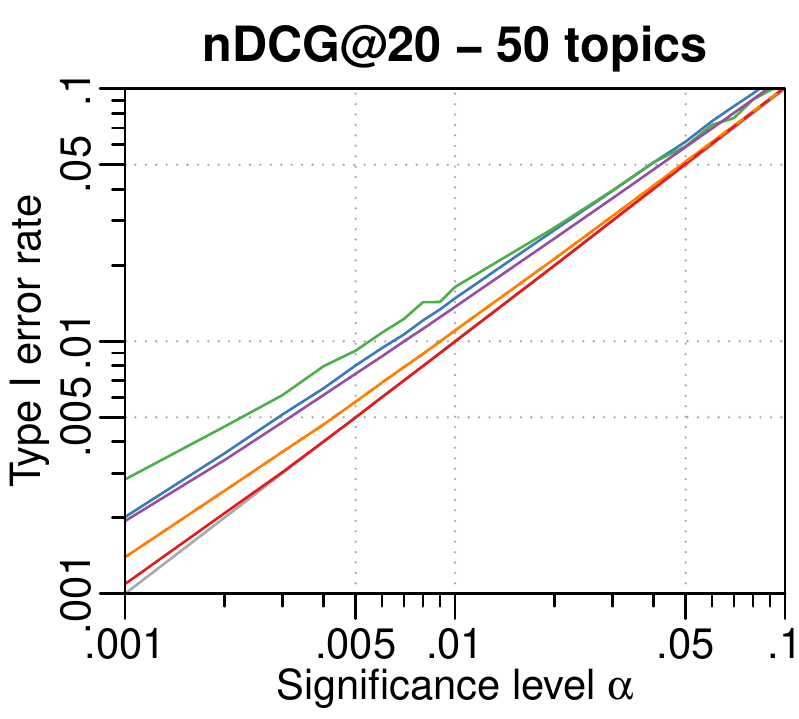}~\includegraphics[scale=\figscale]{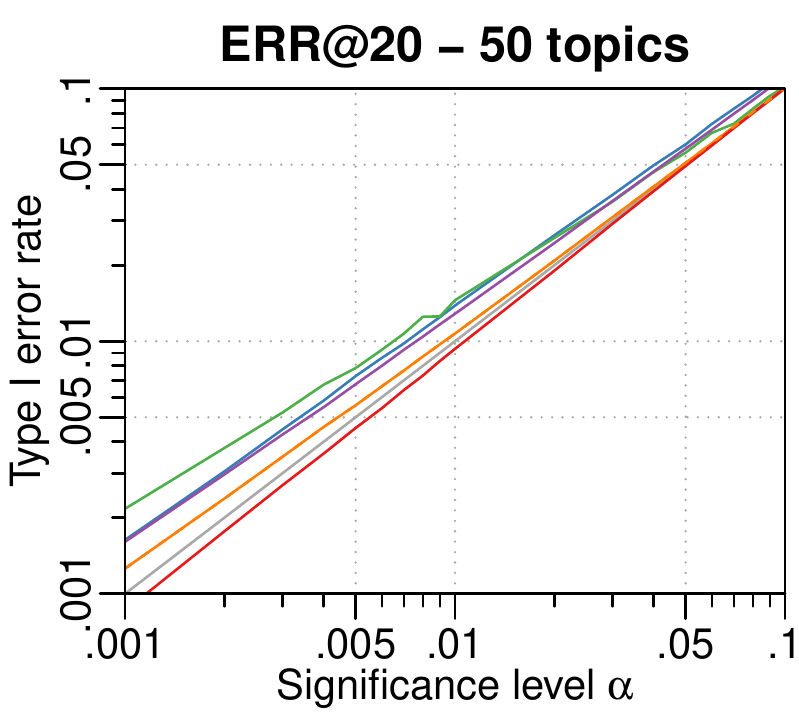}~\includegraphics[scale=\figscale]{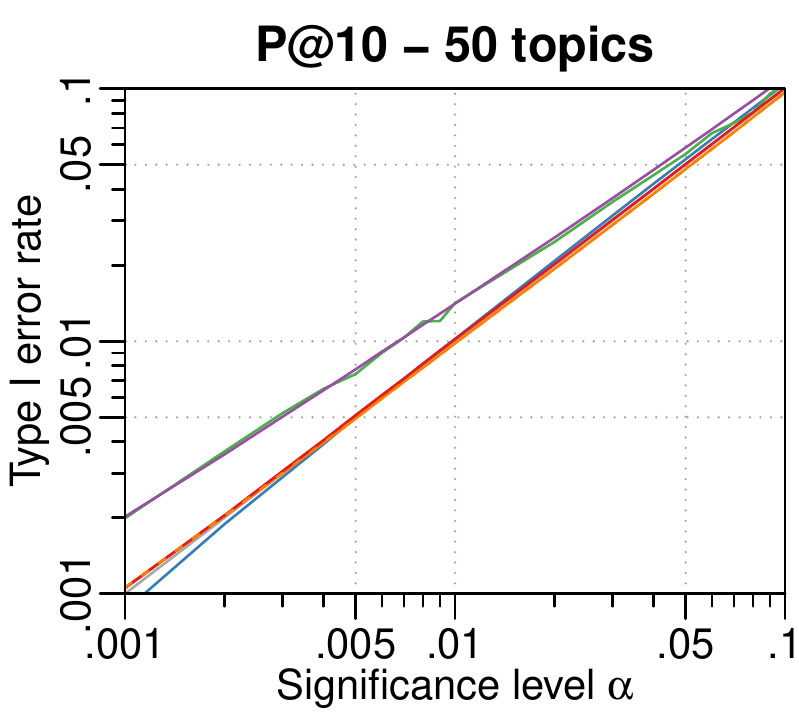}~\includegraphics[scale=\figscale]{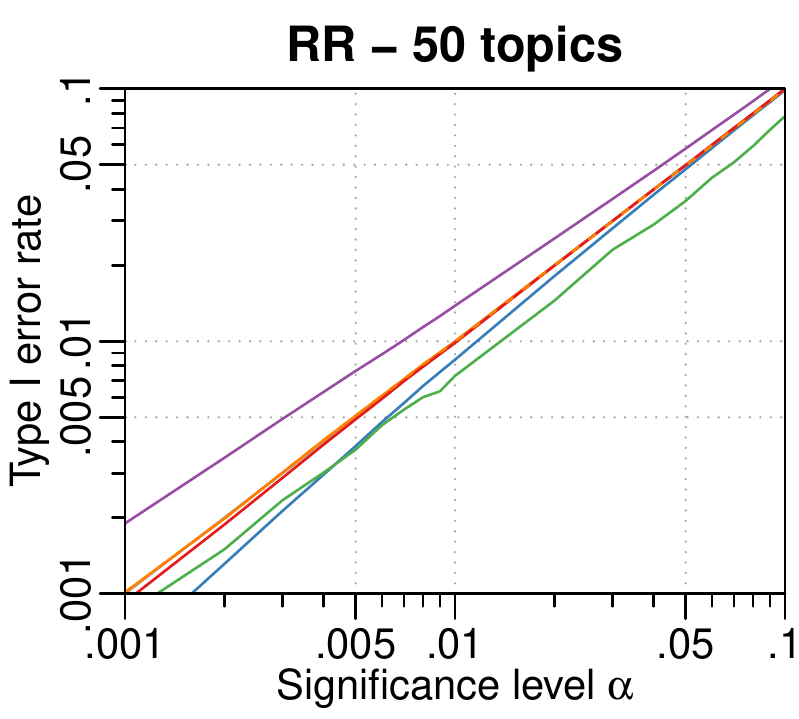}\\
\includegraphics[scale=\figscale]{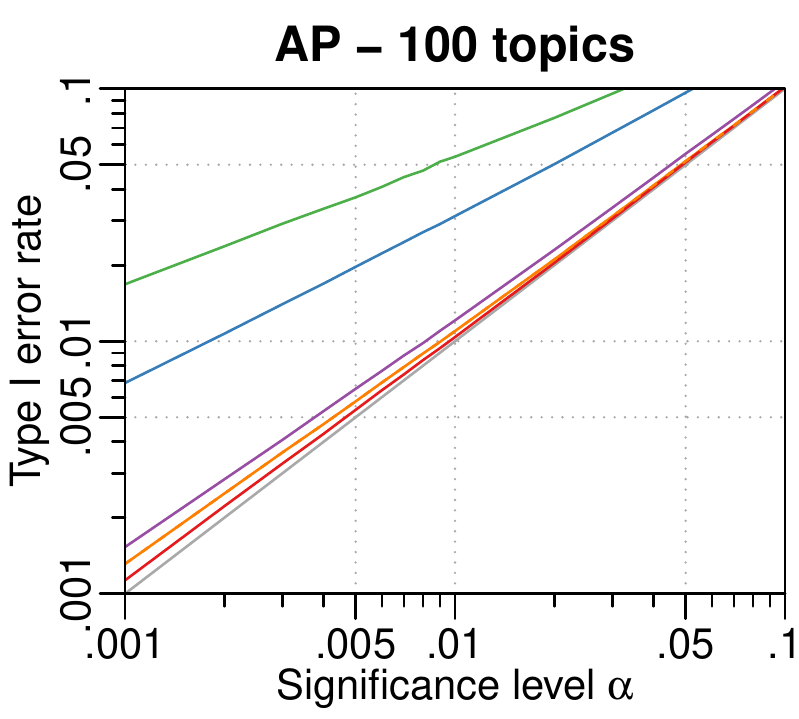}~\includegraphics[scale=\figscale]{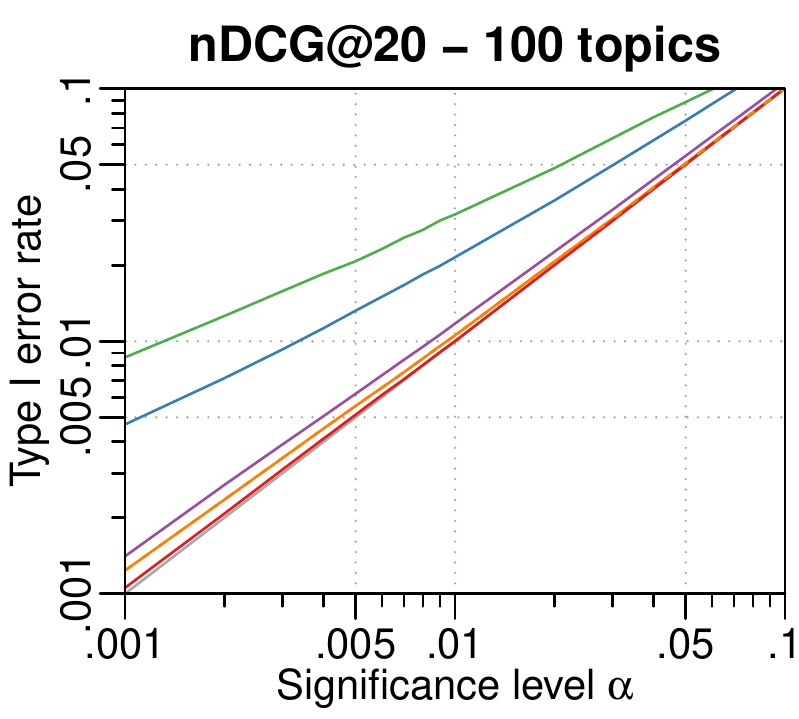}~\includegraphics[scale=\figscale]{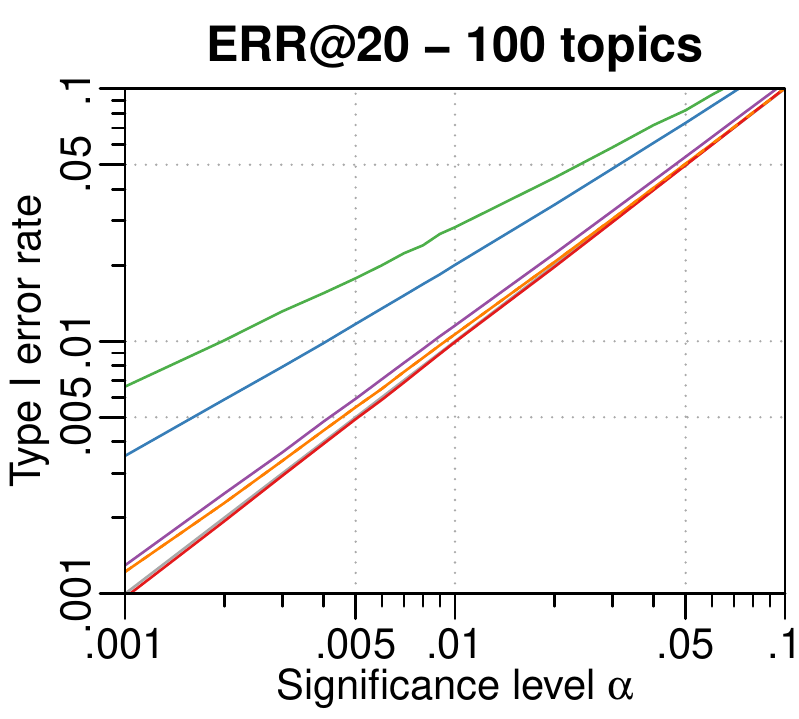}~\includegraphics[scale=\figscale]{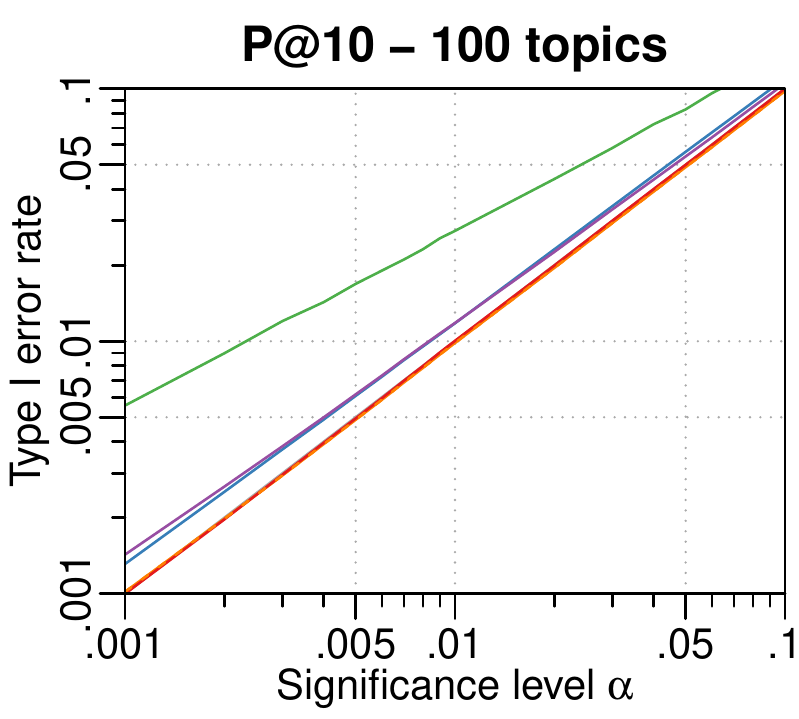}~\includegraphics[scale=\figscale]{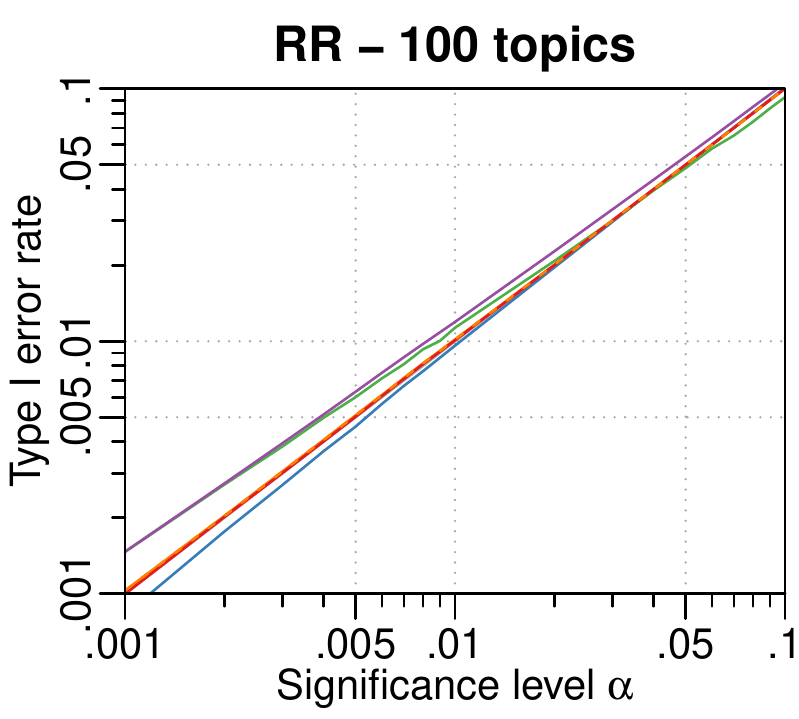}
\caption{Type I error rates of \underline{2-tailed} tests. Plots on the same column correspond to the same effectiveness measure, and plots on the same row correspond to the same topic set size. When indiscernible, the $\bm{t}$, permutation and bootstrap tests overlap.}
\label{fig:type1-2tail}
\end{figure*}

\begin{figure*}[t]
\includegraphics[scale=\figscale]{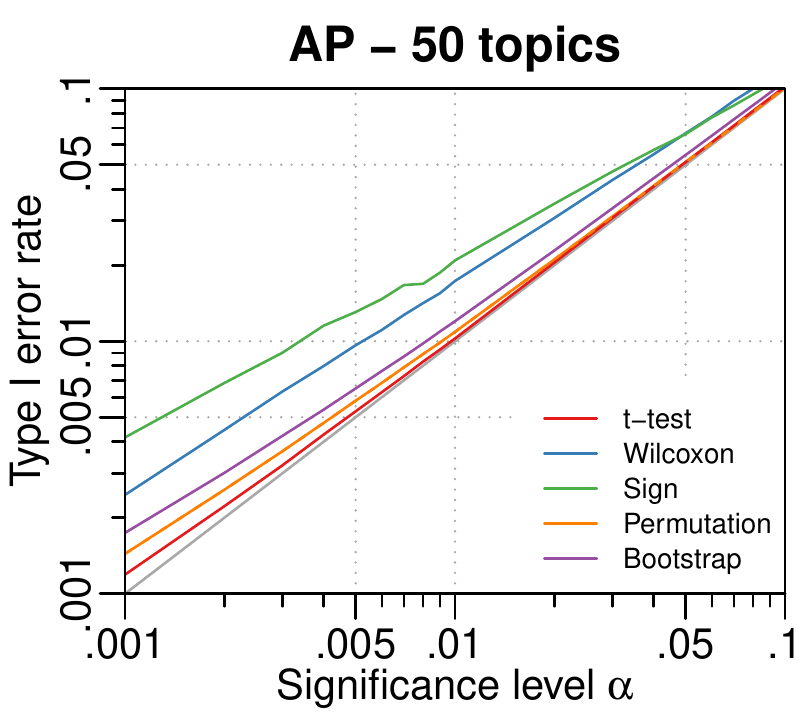}~\includegraphics[scale=\figscale]{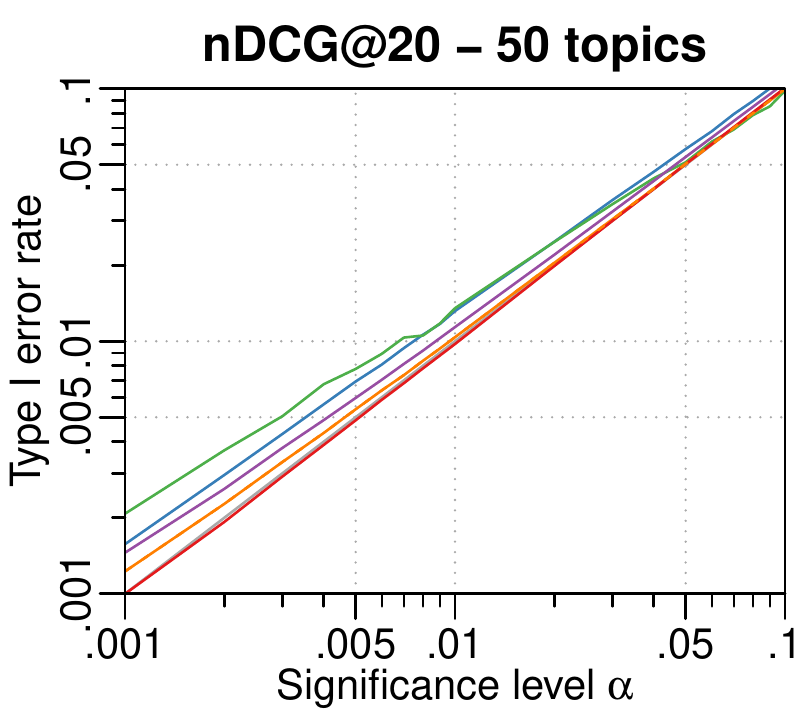}~\includegraphics[scale=\figscale]{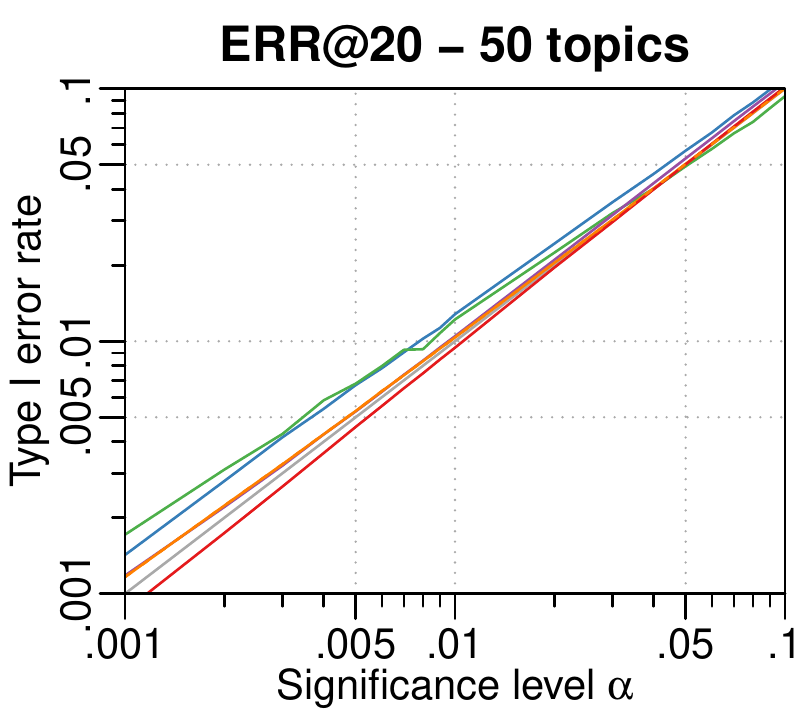}~\includegraphics[scale=\figscale]{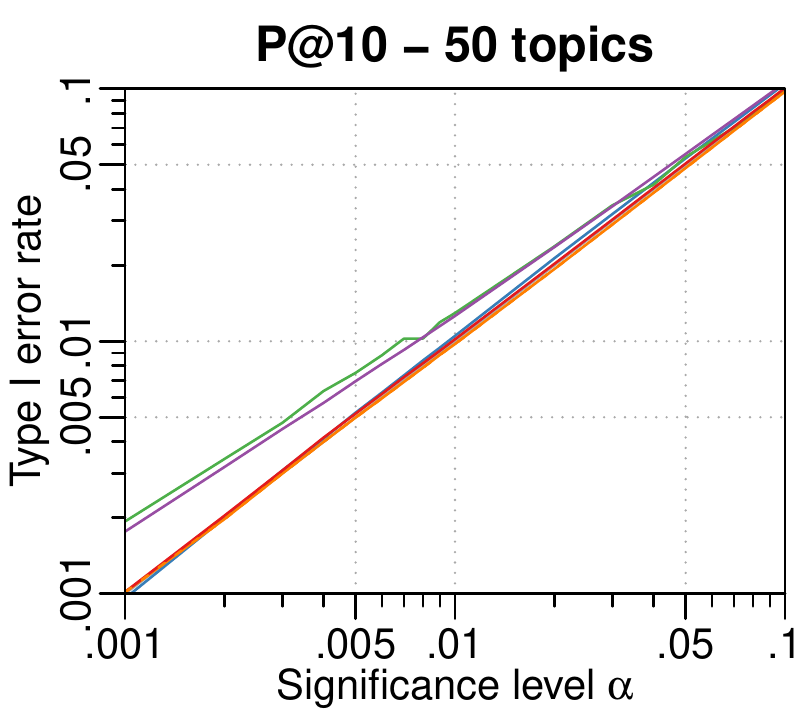}~\includegraphics[scale=\figscale]{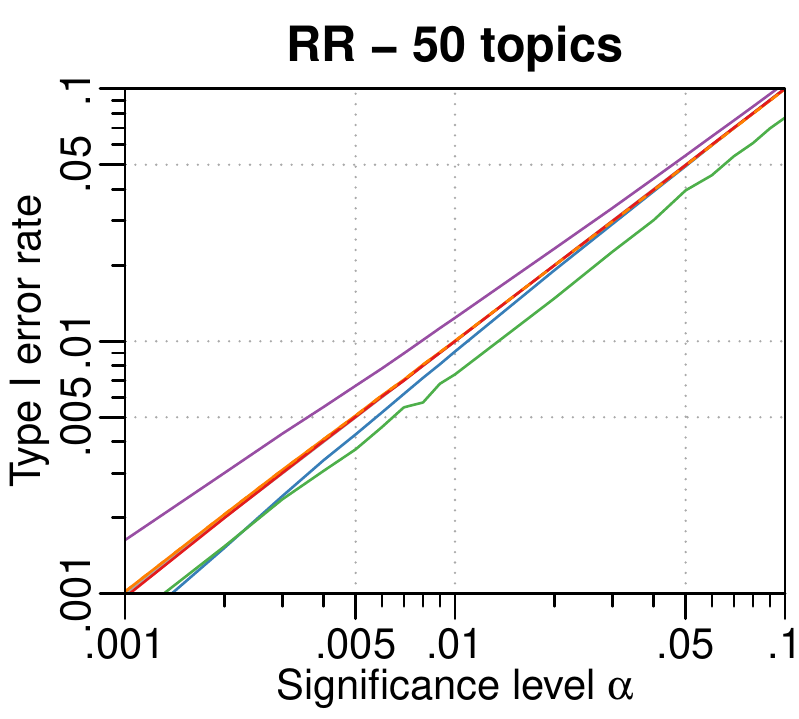}\\
\caption{Type I error rates of \underline{1-tailed} tests with 50 topics. When indiscernible, the $\bm{t}$, permutation and bootstrap tests overlap.}
\label{fig:type1-1tail}
\end{figure*}

In order to evaluate the actual Type I error rate of the statistical tests, we proceed as follows.
For a target measure and topic set size, we randomly select two systems from the same collection and fit the corresponding stochastic model as described in section~\ref{sec:methods:simulation} (same margins). From this model we simulate the scores on $n$ new random topics, and compute the 2-tailed and 1-tailed $p$-values with the 5 tests we consider. Recall that under this stochastic model, both systems have the same distribution and therefore the null hypothesis is known to be true, so any statistically significant result would therefore count as a Type I error.
This is repeated $1,\!667,\!000$ times, leading to $\approx\!8.3$ million 2-tailed $p$-values and $\approx\!8.3$ million 1-tailed $p$-values for every measure and topic set size combination. The grand total is therefore just over $250$ million $p$-values.

Figure~\ref{fig:type1-2tail} shows the actual error rates for the \underline{2-tailed} tests and for significance levels $\alpha$ in $[0.001,0.1]$; each plot summarizes $\approx\!8.3$ million $p$-values. Ideally all lines should go through the diagonal, meaning that the actual error rate ($y$-axis) is the same as the nominal error rate $\alpha$ ($x$-axis). A test going above the diagonal is making more Type I errors than expected, and a test going below the diagonal is a conservative test making fewer errors than expected; both situations should be avoided to the best of our ability. 
Looking at $AP$ scores, we can see that the Wilcoxon and sign tests are consistently making more errors than expected, specially at low $\alpha$ levels. As the sample size increases, they are overconfident and show a clear bias towards small $p$-values. The bootstrap test is similarly overconfident, but it approaches the expected behavior as the sample size increases, because the sampling distribution is better estimated with large samples. The permutation test behaves quite better and also approaches the diagonal as sample size increases, and the $t$-test is remarkably close to ideal behavior even with small sample sizes. With $nDCG@20$ and $ERR@20$ we see very similar behavior, with the $t$-test tracking the ideal error rate nearly perfectly, perhaps with the exception of the $t$-test being conservative at low $\alpha$ levels with small samples in $ERR@20$.

\begin{figure*}[!ht]
\includegraphics[scale=\figscale]{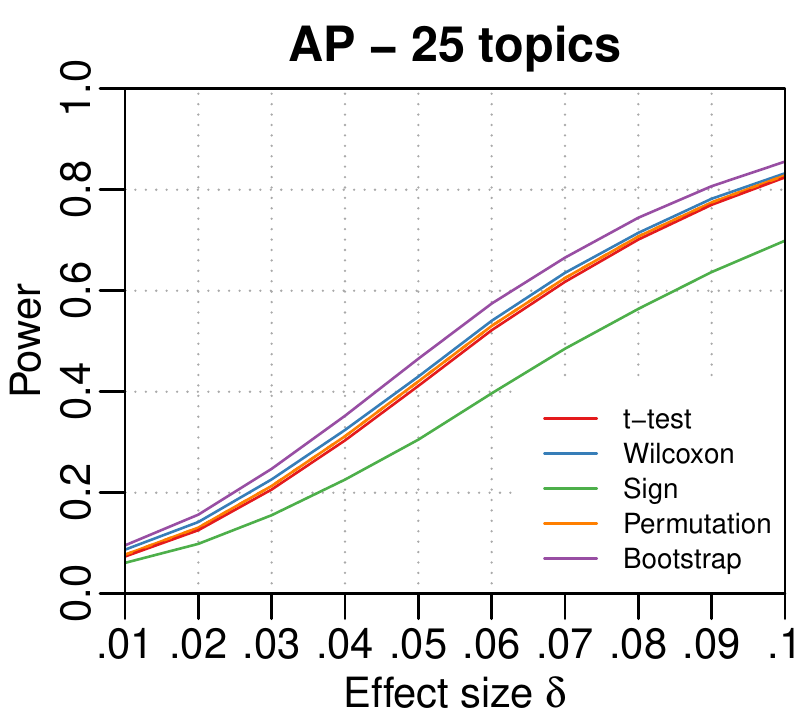}~\includegraphics[scale=\figscale]{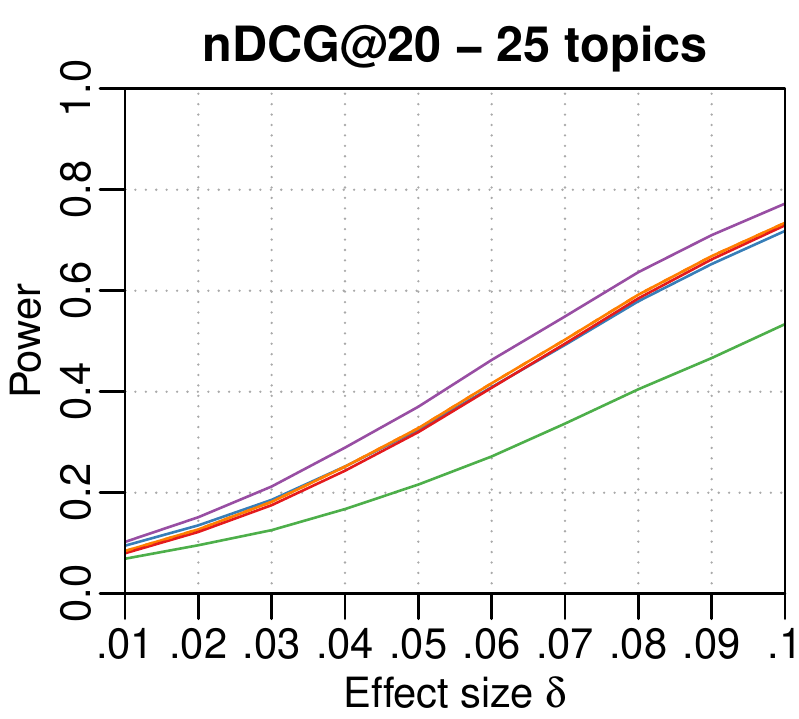}~\includegraphics[scale=\figscale]{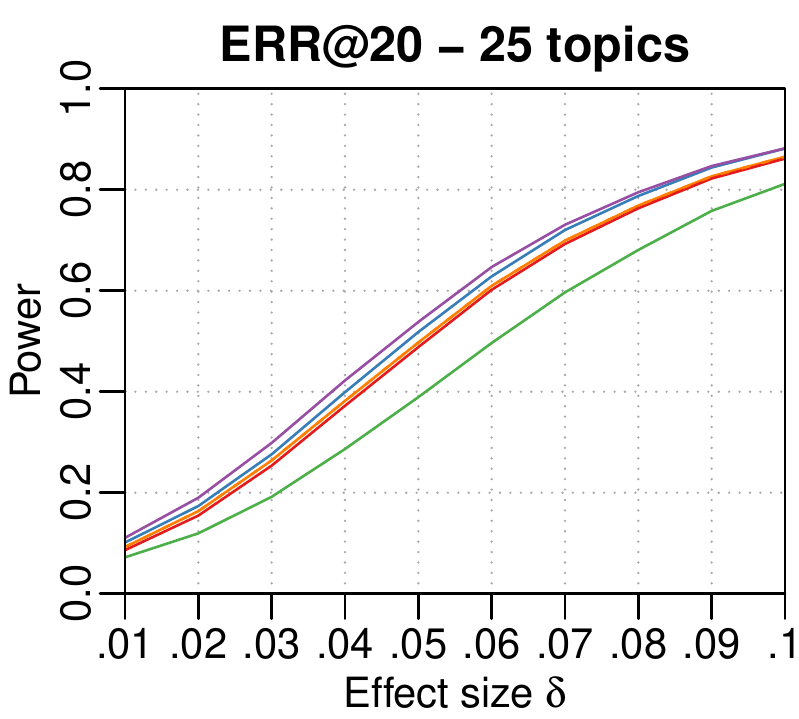}~\includegraphics[scale=\figscale]{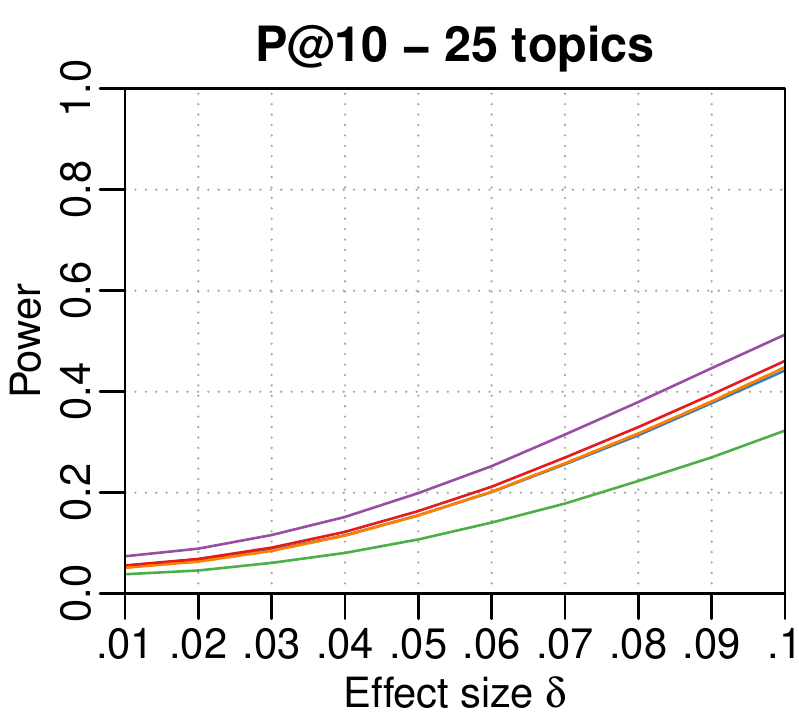}~\includegraphics[scale=\figscale]{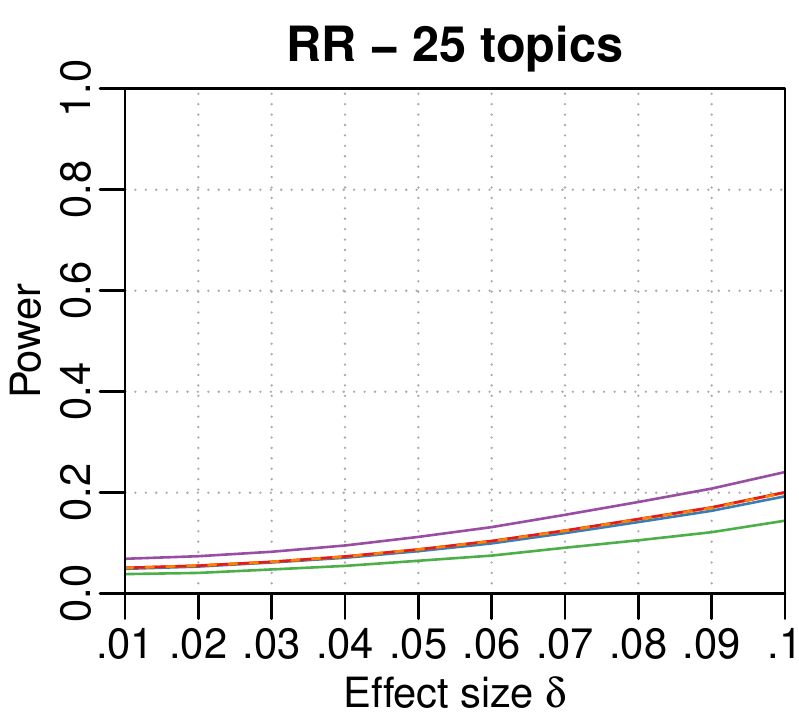}\\
\includegraphics[scale=\figscale]{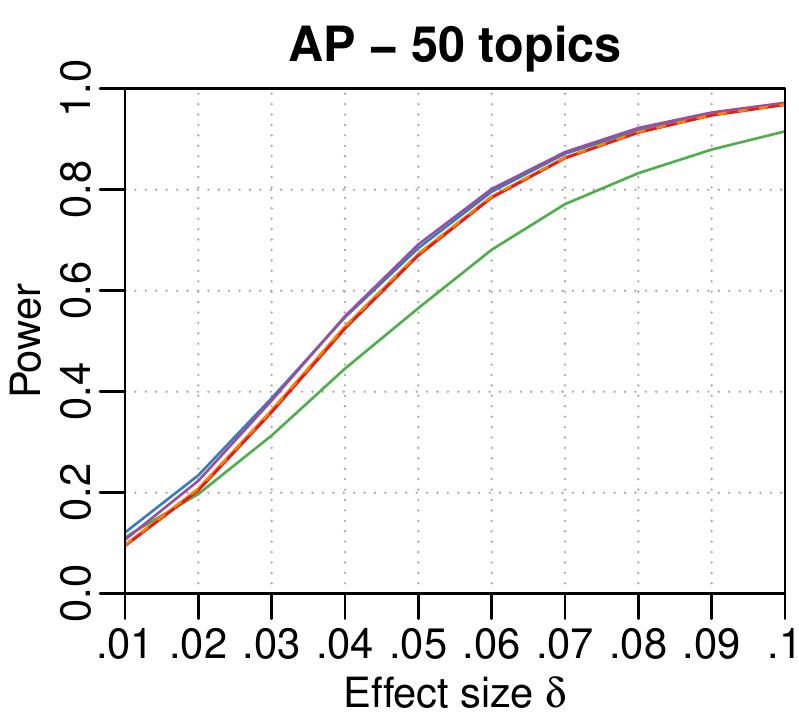}~\includegraphics[scale=\figscale]{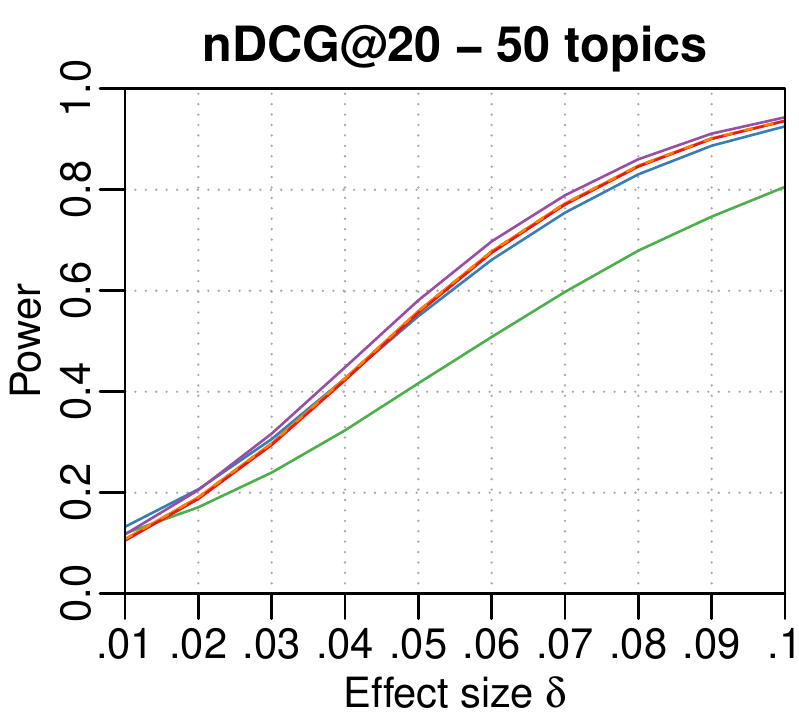}~\includegraphics[scale=\figscale]{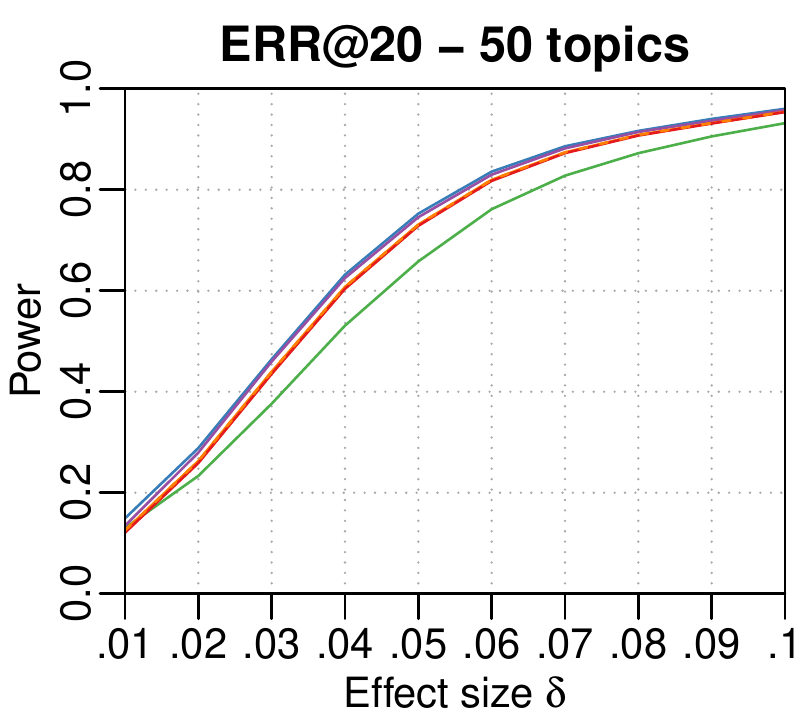}~\includegraphics[scale=\figscale]{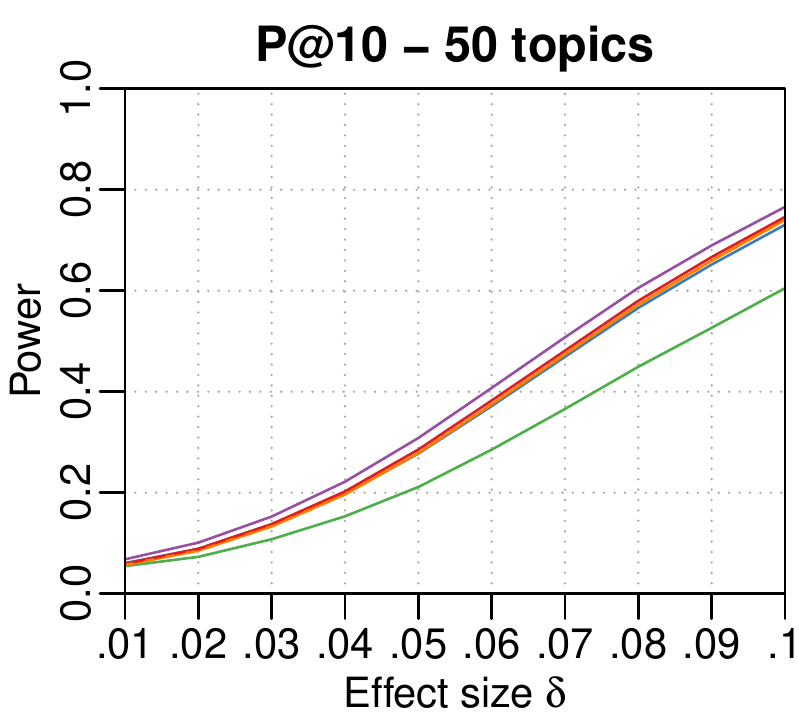}~\includegraphics[scale=\figscale]{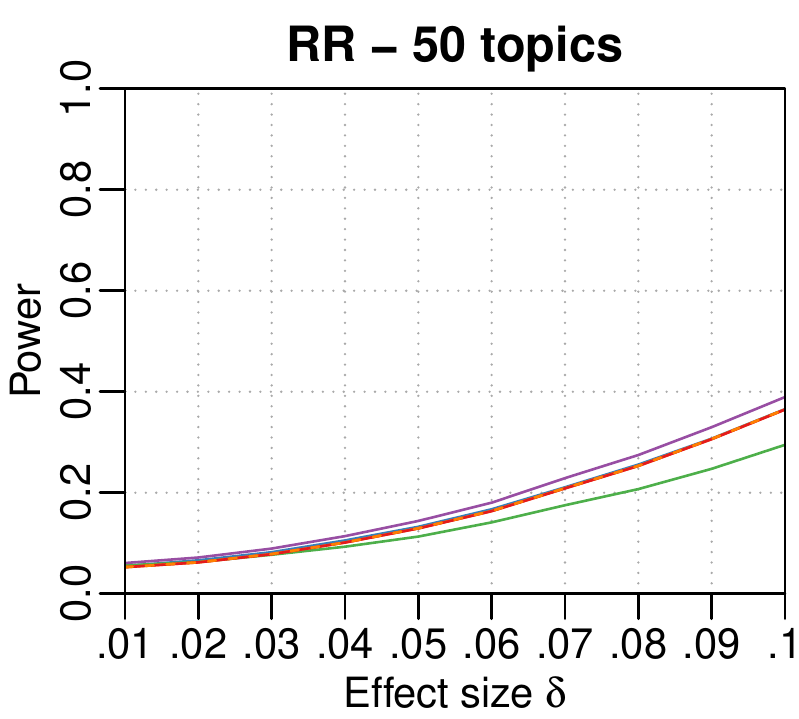}\\
\includegraphics[scale=\figscale]{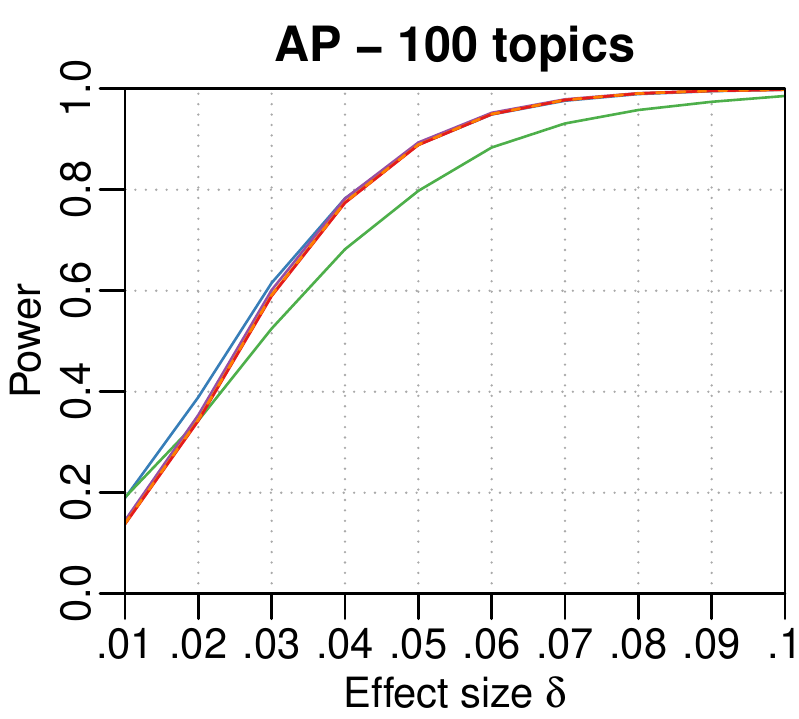}~\includegraphics[scale=\figscale]{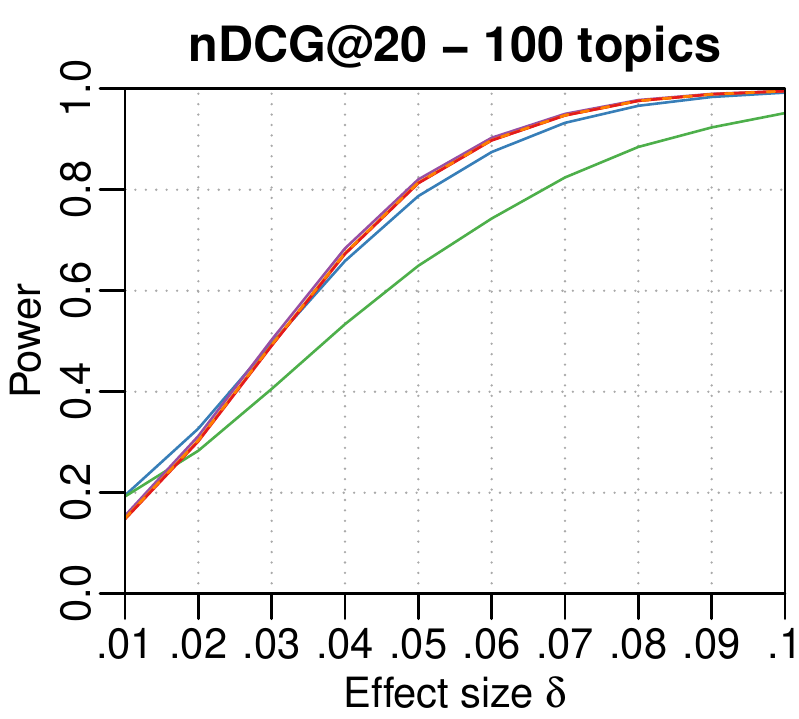}~\includegraphics[scale=\figscale]{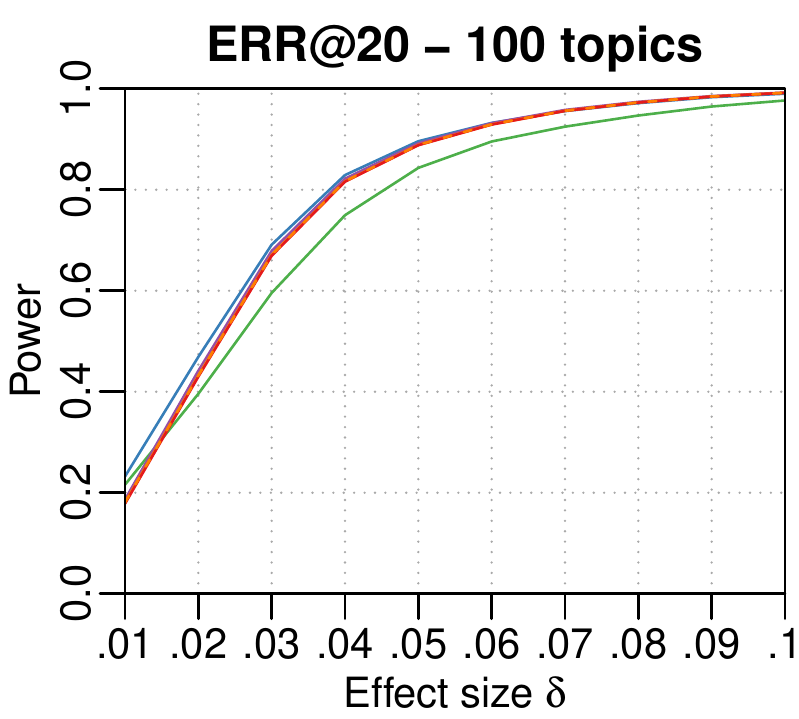}~\includegraphics[scale=\figscale]{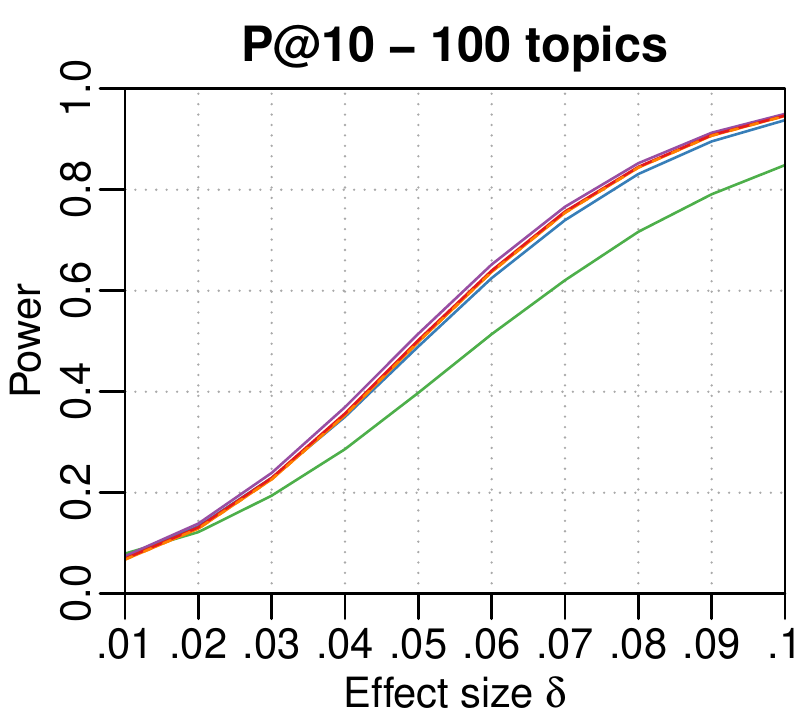}~\includegraphics[scale=\figscale]{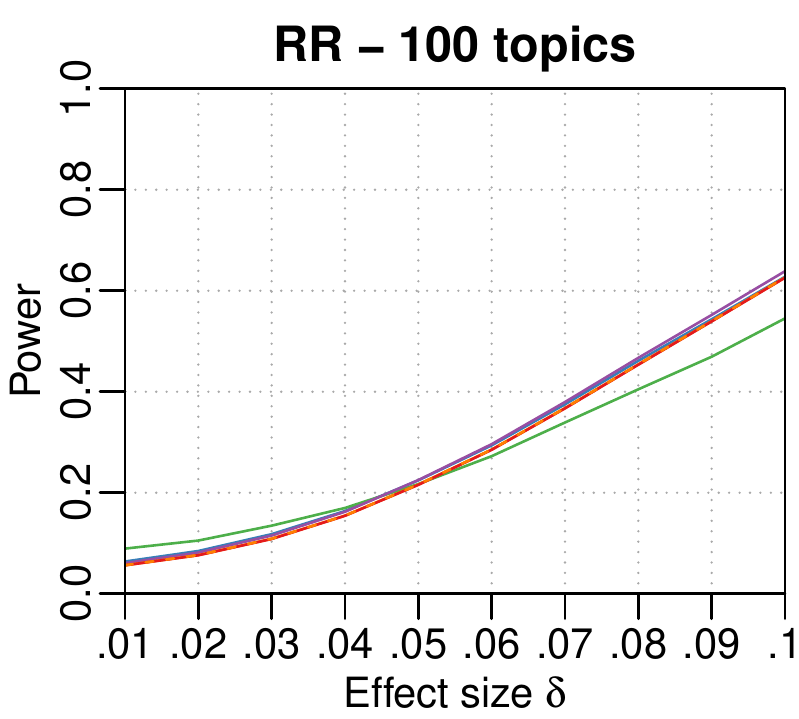}
\caption{Power of \underline{2-tailed} tests as a function of $\bm{\delta}$ and at $\bm{\alpha=0.05}$. Plots on the same column correspond to the same topic set size, and plots on the same row correspond to the same effectiveness measure.}
\label{fig:power-2tail}
\end{figure*}

When we look at a measure like $P@10$ that produces clearly discrete distributions, it is remarkable how well the $t$ and permutation tests behave. This is probably because the distributions of differences in $P@10$ are very symmetric in practice, which facilitates the effect of the Central Limit Theorem. For the same reason, the Wilcoxon test behaves better than before, but again becomes overconfident with large samples. The sign test also suffers from large sample sizes by increasing error rates, while the bootstrap test behaves better with more topics. For $RR$ we see very similar results, again with the exception of the $t$-test being conservative at low $\alpha$ levels with small samples. The permutation test shows nearly ideal behavior for $RR$ too.

Figure~\ref{fig:type1-1tail} shows similar plots but for the \underline{1-tailed} case and only with $n=50$ topics. The results are consistent with the 2-tailed case: the permutation and $t$-tests show nearly ideal behavior and the bootstrap test has a bias towards small $p$-values. When sample size increases (not shown\footnote{Plots for other topic set sizes are provided as supplementary material.}), this bias is reduced, but the Wilcoxon and sign tests become even more unreliable again.

For the typical IR case of $n=50$ topics and $\alpha=0.05$, we see that the $t$-test and the permutation tests maintain a Type I error rate of just 0.05, while the bootstrap test is at 0.059 in the 2-tailed case and 0.054 in the 1-tailed case. At the $\alpha=0.01$ level, the $t$-test and the permutation test maintain a 0.01 error rate, with the bootstrap going up to 0.014.

\subsection{Type II Errors}\label{sec:results:type2}

In order to evaluate the actual Type II error rate, we need to simulate from systems whose true mean scores are at a distance $\delta$ from each other, and then observe how error rates are affected by the magnitude of this distance. We first randomly select a baseline system \sysB from the bottom 75\% of runs for a collection and measure, and then pre-select the 10 systems whose mean score is closest to the target $\mu_B+\delta$. From these 10, we randomly choose the experimental system \sysE and fit the stochastic model to simulate, but we additionally transform $F_E$ so that its expected value is $\mu_E=\mu_B+\delta$, as described in section~\ref{sec:methods:simulation}. Under this model, the null hypothesis is known to be false, so any result that is not statistically significant would therefore count as a Type II error. This is repeated $167,\!000$ times for each value of $\delta$ in $0.01, 0.02, \dots, 0.1$. In total, this leads to $8.35$ million 2-tailed $p$-values and $8.35$ million 1-tailed $p$-values for every measure and topic set size combination. The grand total is therefore just over 250 million $p$-values as well.

\begin{figure*}[!ht]
\includegraphics[scale=\figscale]{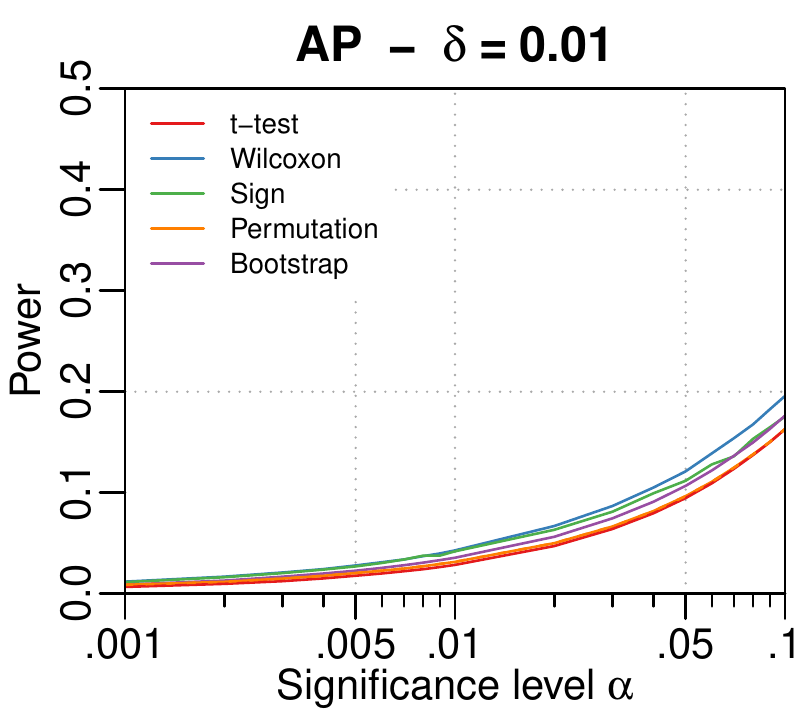}~\includegraphics[scale=\figscale]{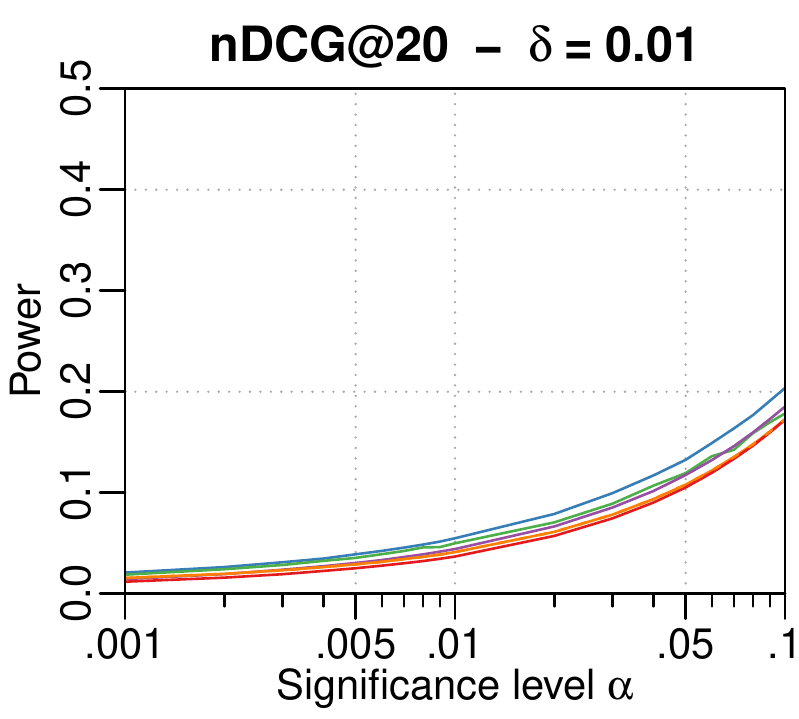}~\includegraphics[scale=\figscale]{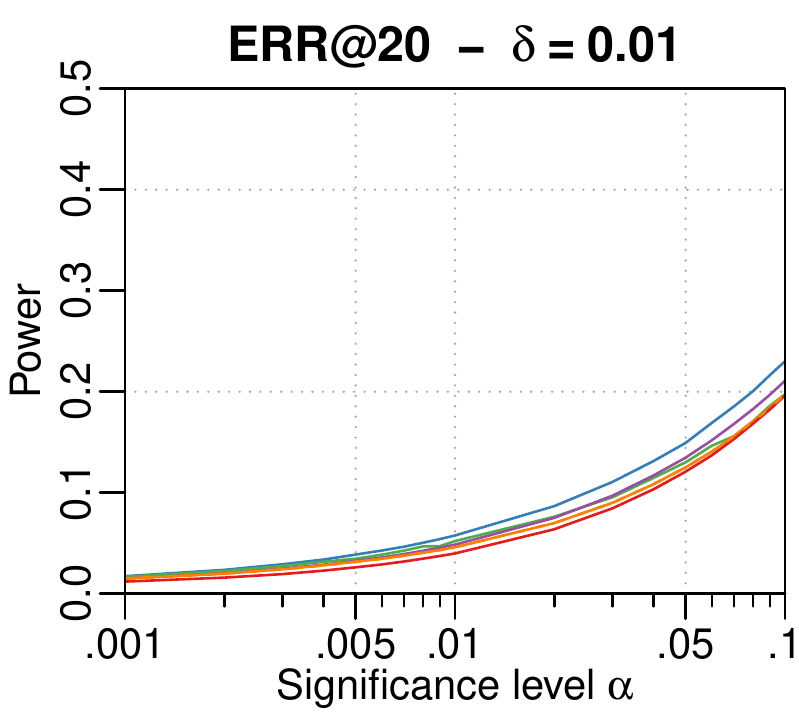}~\includegraphics[scale=\figscale]{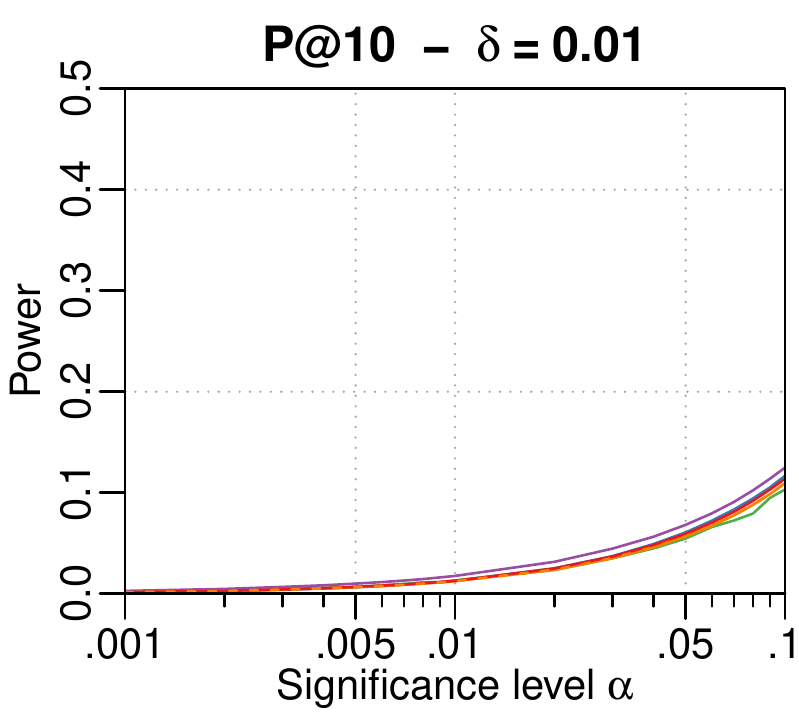}~\includegraphics[scale=\figscale]{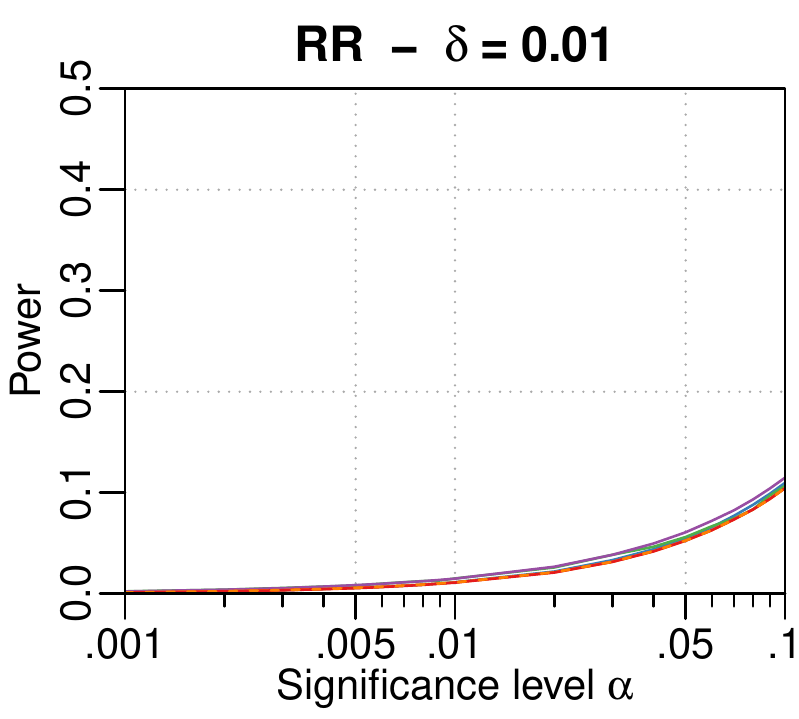}\\
\includegraphics[scale=\figscale]{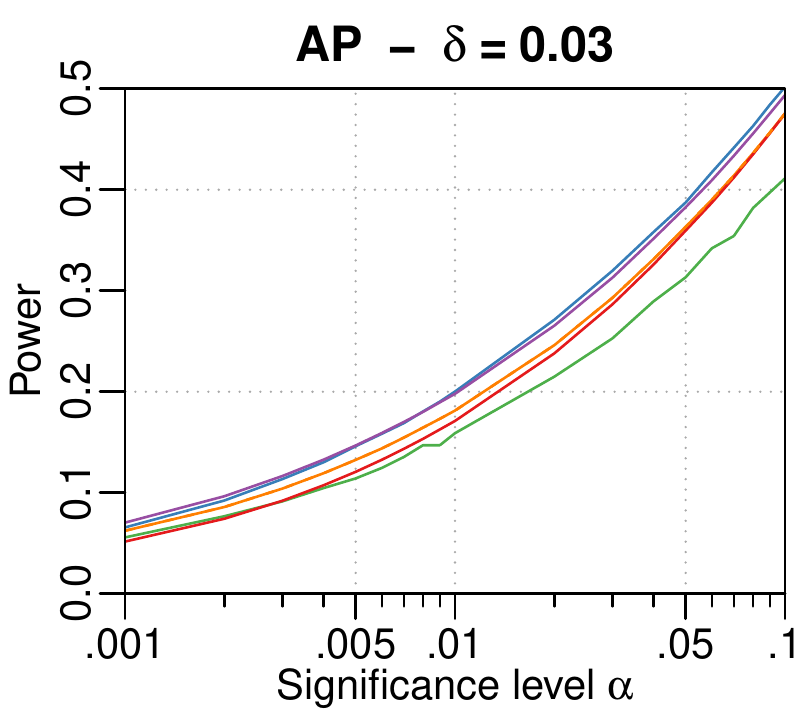}~\includegraphics[scale=\figscale]{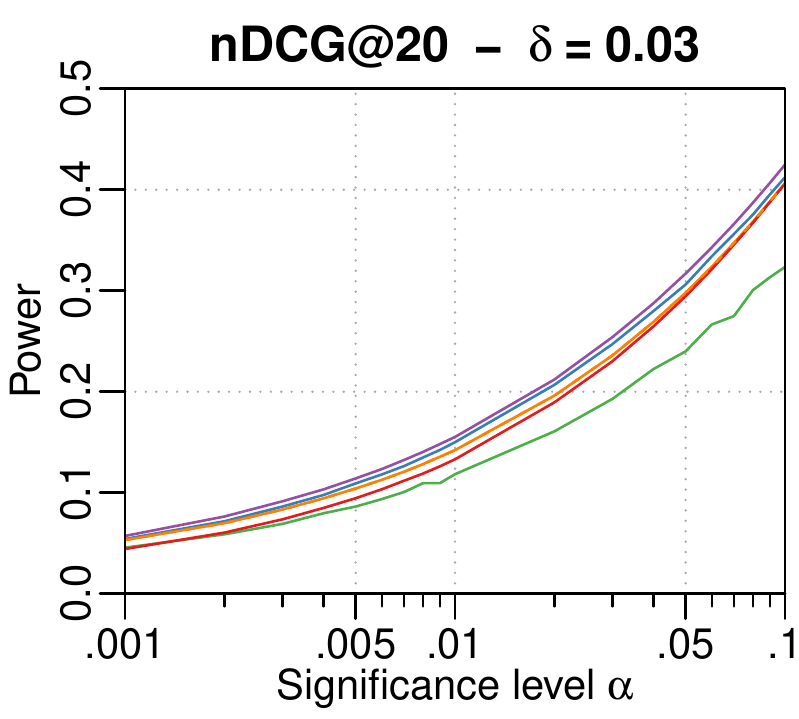}~\includegraphics[scale=\figscale]{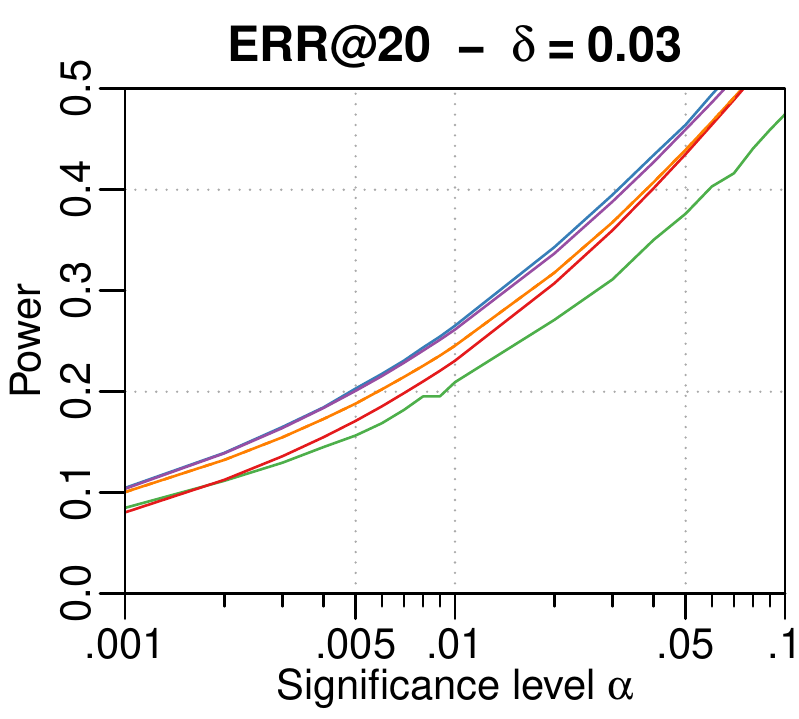}~\includegraphics[scale=\figscale]{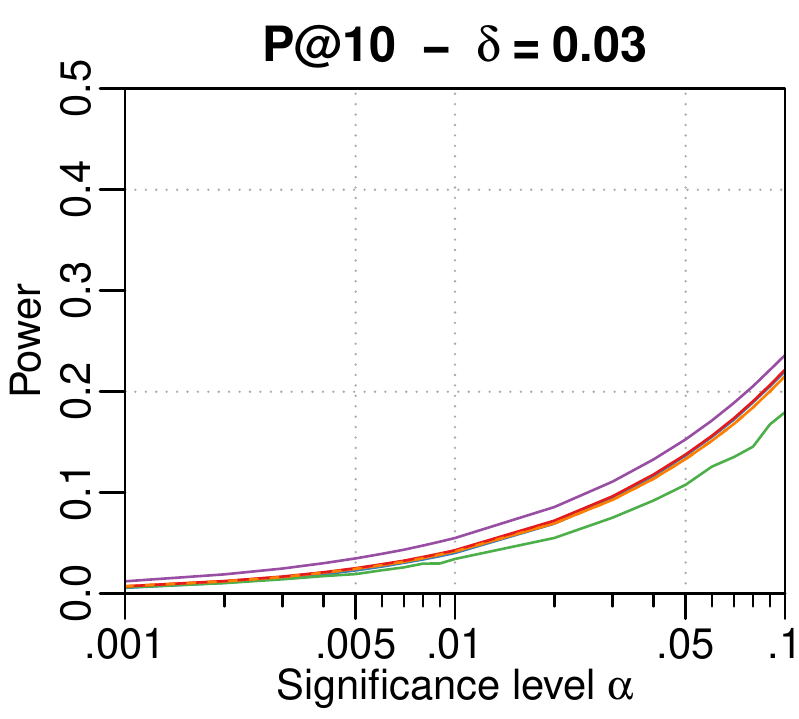}~\includegraphics[scale=\figscale]{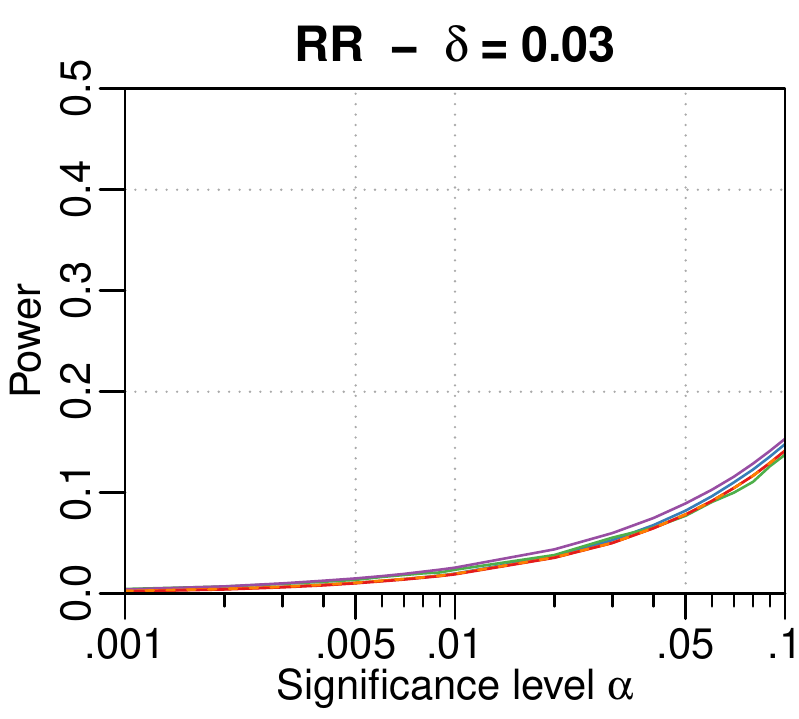}\\
\caption{Power of \underline{2-tailed} tests as a function of $\bm{\alpha}$ and at $\bm{\delta=0.01,0.03}$, with  50 topics.}
\label{fig:power-2tail-alpha}
\end{figure*}

As is customary in this kind of analysis, we report power curves instead of Type II error rates, and do so only for the typical significance level $\alpha=0.05$.\footnote{Plots for other $\alpha$ levels are provided online as supplementary material.} Figure~\ref{fig:power-2tail} shows the power curves of the \underline{2-tailed} tests as a function of the effect size $\delta$. The first thing we notice is that there are clear differences across measures; this is due to the variability of their distributions. For instance, differences in $P@10$ or $RR$ have between 2 and 4 times the standard deviation of the other measures. Across all measures we can also see clearly how power increases with the sample size and with the effect size. These plots clearly visualize the effect of the three main factors that affect statistical power: variability, effect size and sample size. The definition of the $t$ statistic in eq.~\eqref{eq:t} nicely formulates this.

We can see nearly the same pattern across measures. The sign test is consistently less powerful because it does not take magnitudes into account, and the bootstrap test is nearly consistently the most powerful, specially with small samples. The second place is disputed between the Wilcoxon, $t$ and permutation tests. For instance, the Wilcoxon test is slightly more powerful for $AP$ or $ERR@20$, but less so for $nDCG@20$ or $P@10$. For the most part though, these three tests perform very similarly, specially the $t$ and permutation tests. When the sample size is large, all tests are nearly indistinguishable in terms of power, with the clear exception of the sign test. Results in the \underline{1-tailed} case are virtually the same, except that power is of course higher. The reader is referred to the online supplementary material for the full set of results.

Figure~\ref{fig:power-2tail-alpha} offers a different perspective by plotting power as a function of the significance level, and for the selection $\delta\!=\!0.01, 0.03$ and 50 topics. These plots confirm that the bootstrap and Wilcoxon tests have the highest power, specially for $AP$, $nDCG@20$ and $ERR@20$. That the bootstrap-shift test appears to be more powerful was also noted for instance by~\citet{smucker2007comparison,urbano2013comparison}, but the slightly higher power of the Wilcoxon test comes as a surprise to these authors. \citet[\S5.7,\S5.11]{conover1999practical} points out that, for certain heavy-tailed distributions, the Wilcoxon test may indeed have higher asymptotic relative efficiency compared to the $t$ and permutation tests. For small samples,~\citet{conover1978asymptotic} report similar findings in the two-sample case.

In summary, all tests except the sign test behave very similarly, with very small differences in practice. The bootstrap and Wilcoxon tests are consistently a little more powerful, but the results from last section indicate that they are also more likely to make Type I errors. Given that the $t$-test and the permutation test behave almost ideally in terms of Type I errors, and similar to the others in terms of power, it seems clear that they are the best choice, also consistently across measures and sample sizes. 

\begin{figure*}[t]
\includegraphics[scale=\figscale]{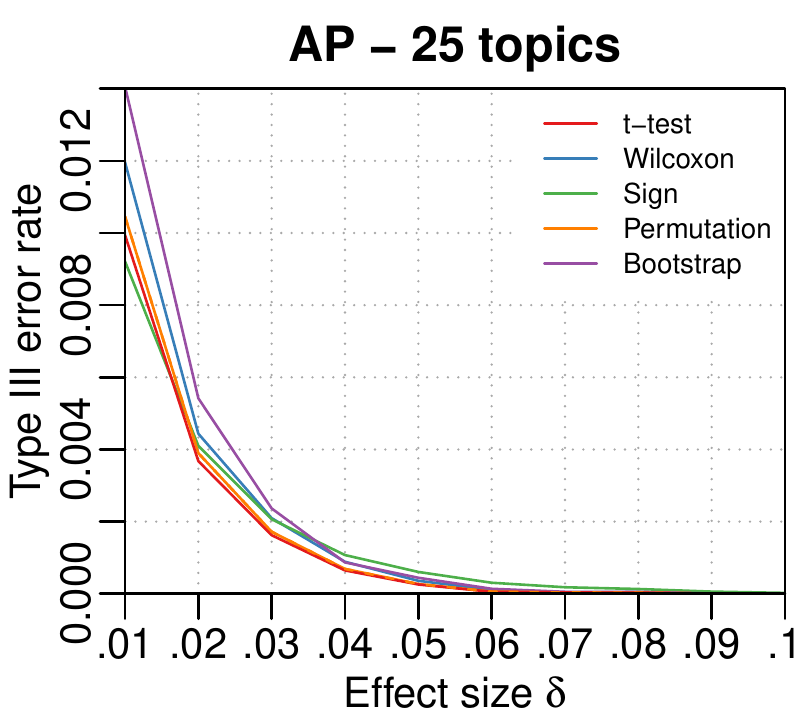}~\includegraphics[scale=\figscale]{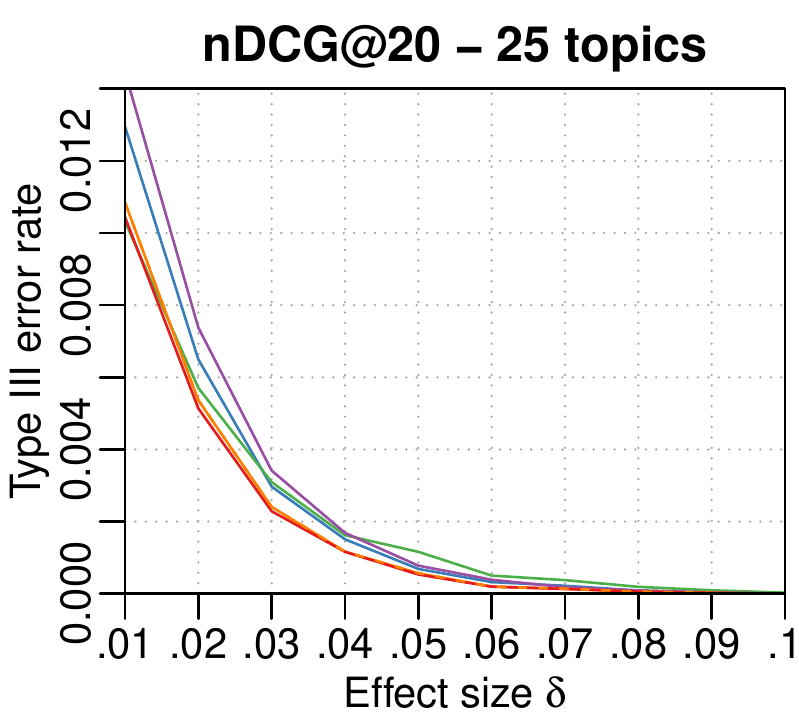}~\includegraphics[scale=\figscale]{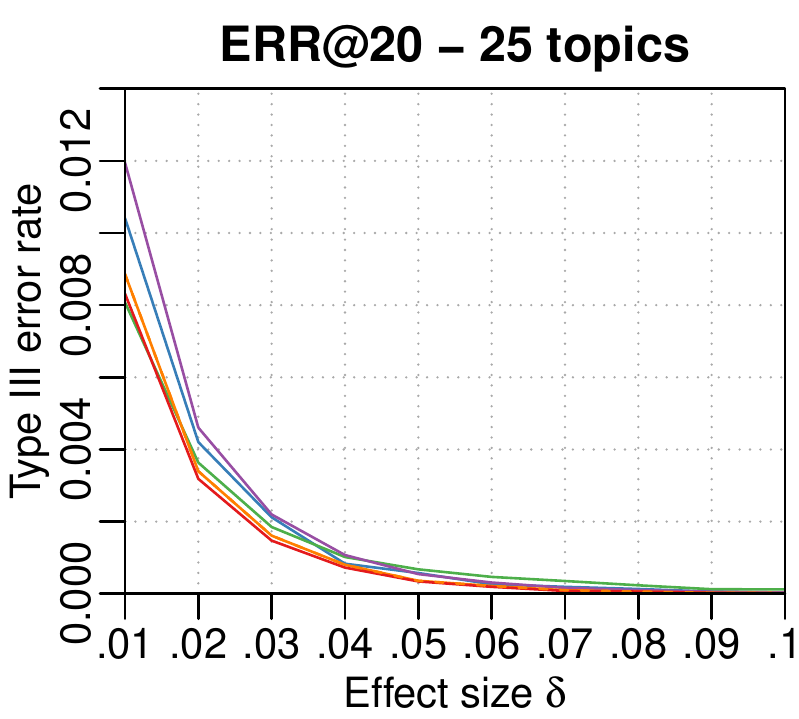}~\includegraphics[scale=\figscale]{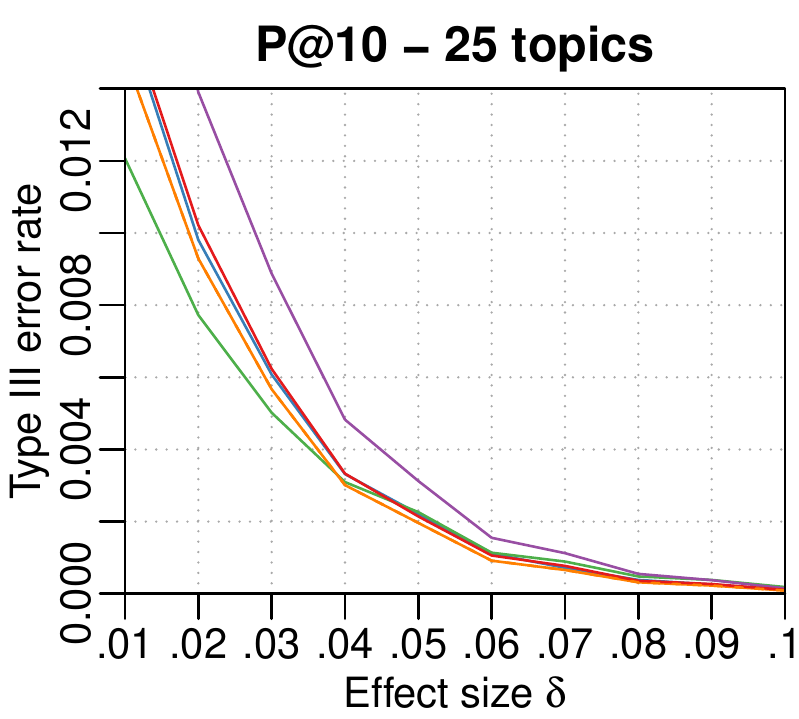}~\includegraphics[scale=\figscale]{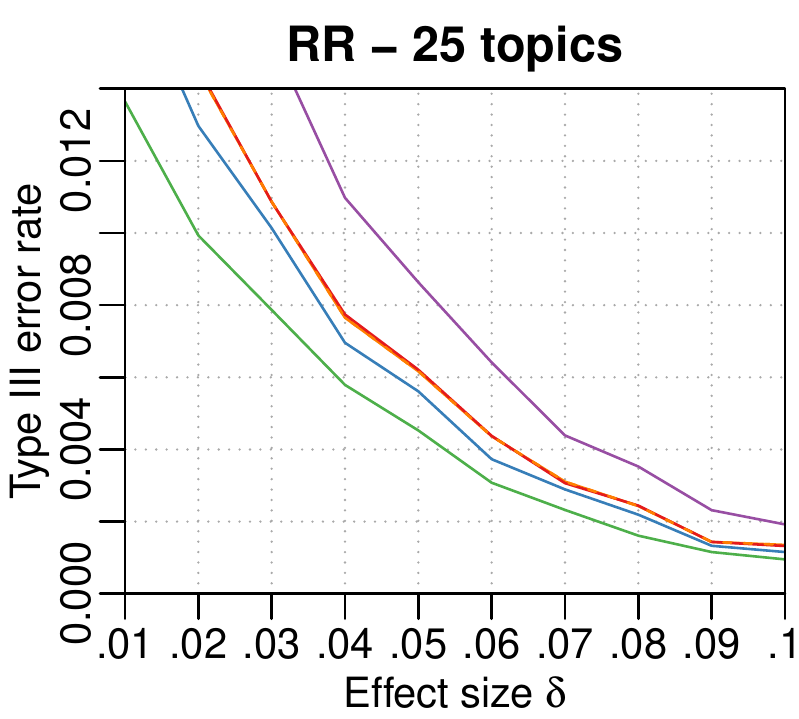}\\
\includegraphics[scale=\figscale]{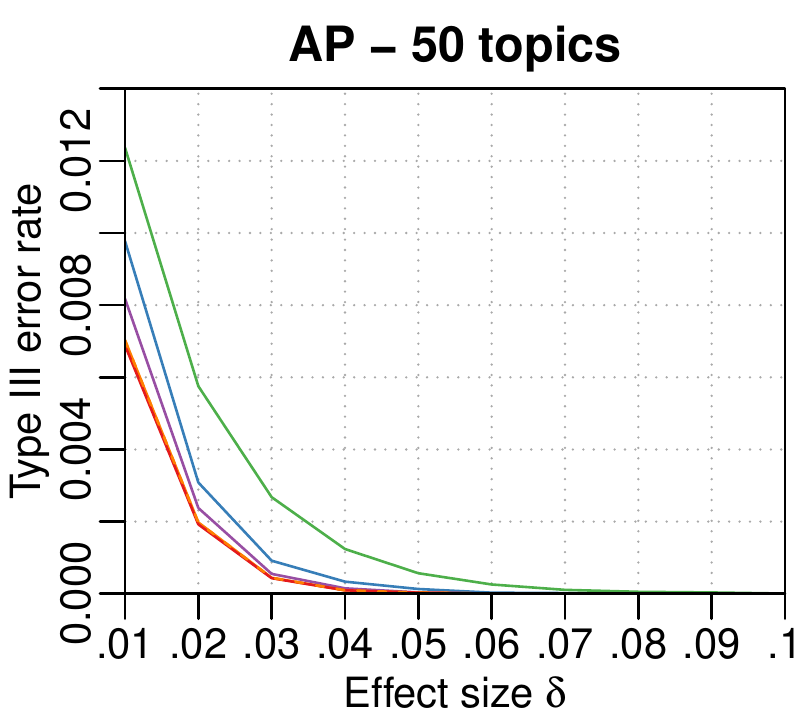}~\includegraphics[scale=\figscale]{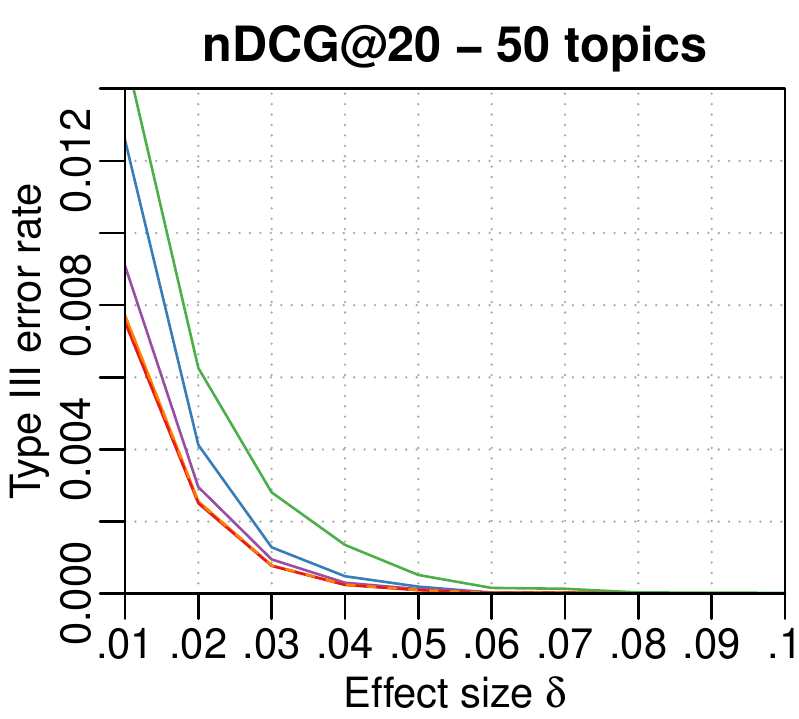}~\includegraphics[scale=\figscale]{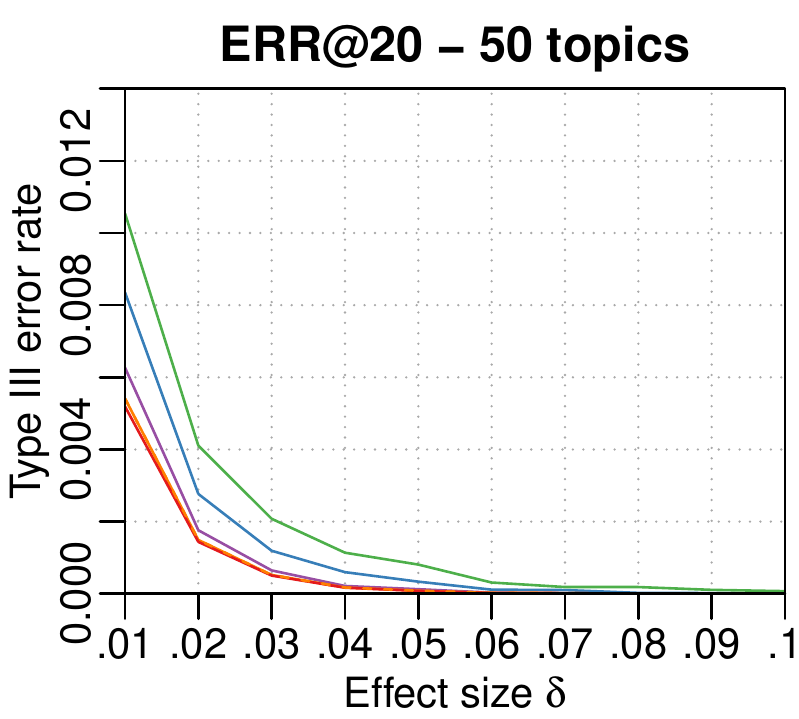}~\includegraphics[scale=\figscale]{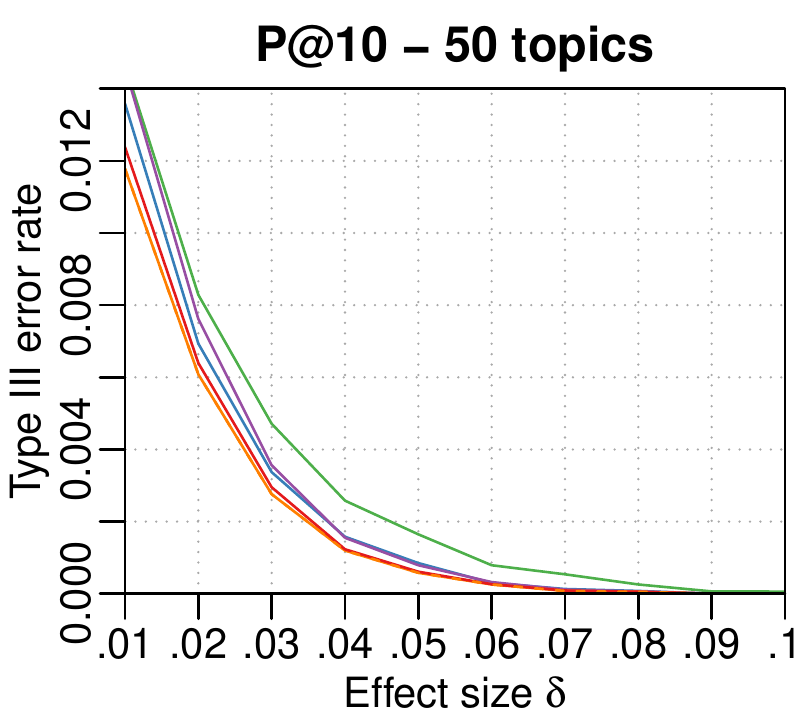}~\includegraphics[scale=\figscale]{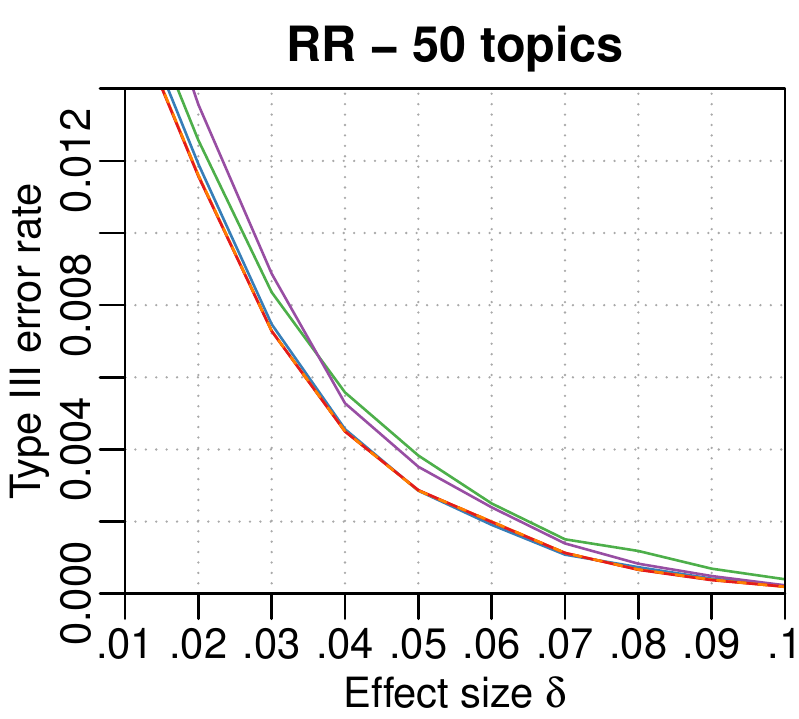}\\
\includegraphics[scale=\figscale]{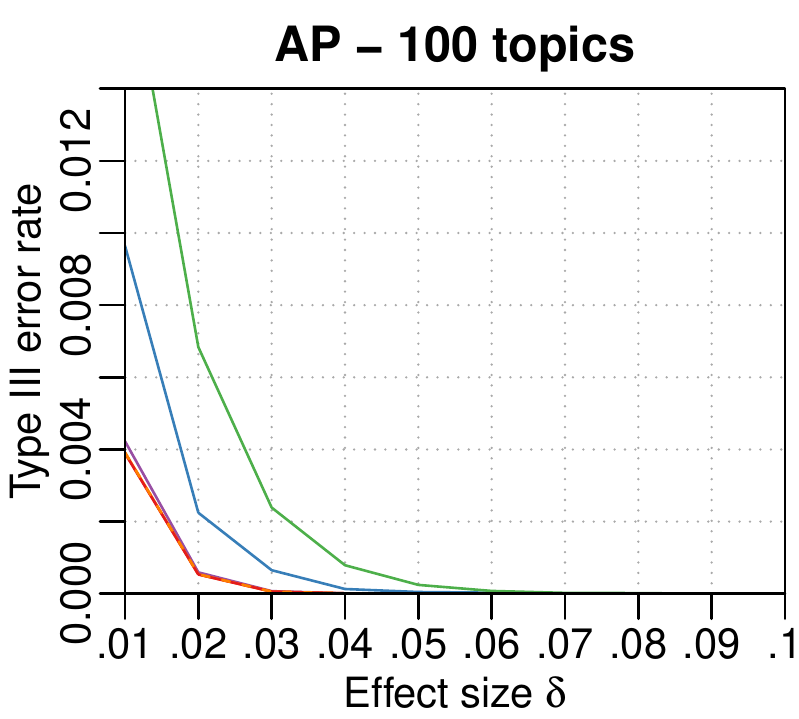}~\includegraphics[scale=\figscale]{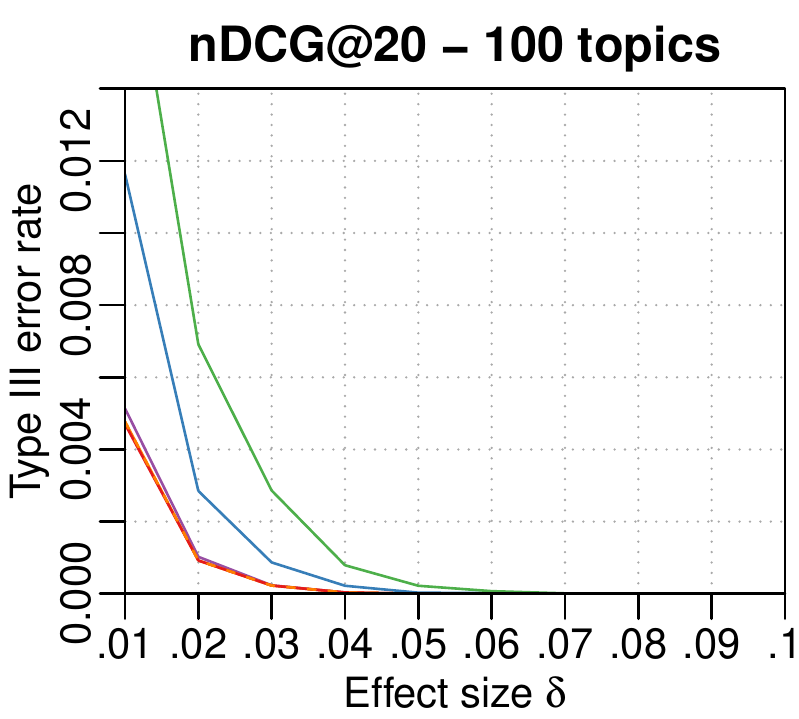}~\includegraphics[scale=\figscale]{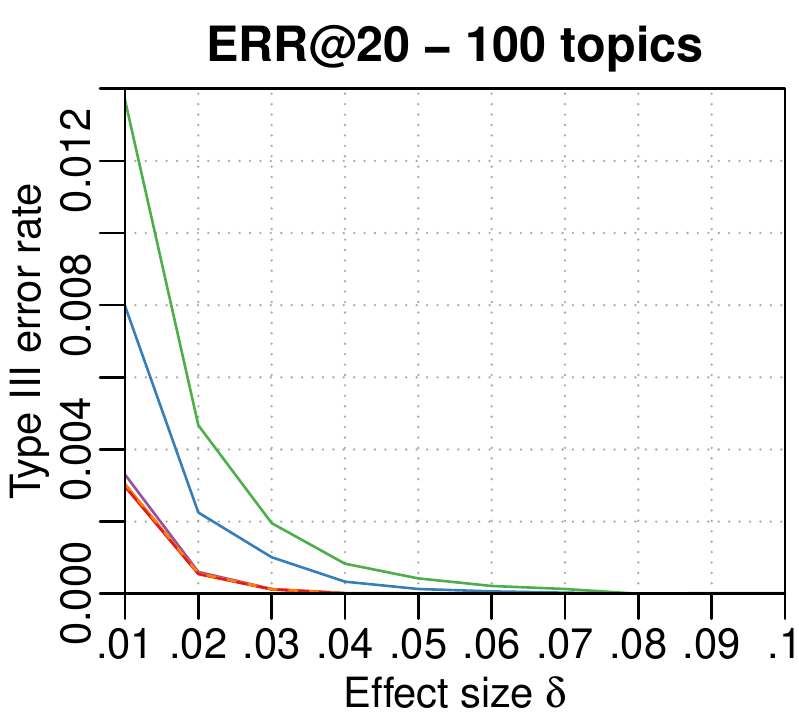}~\includegraphics[scale=\figscale]{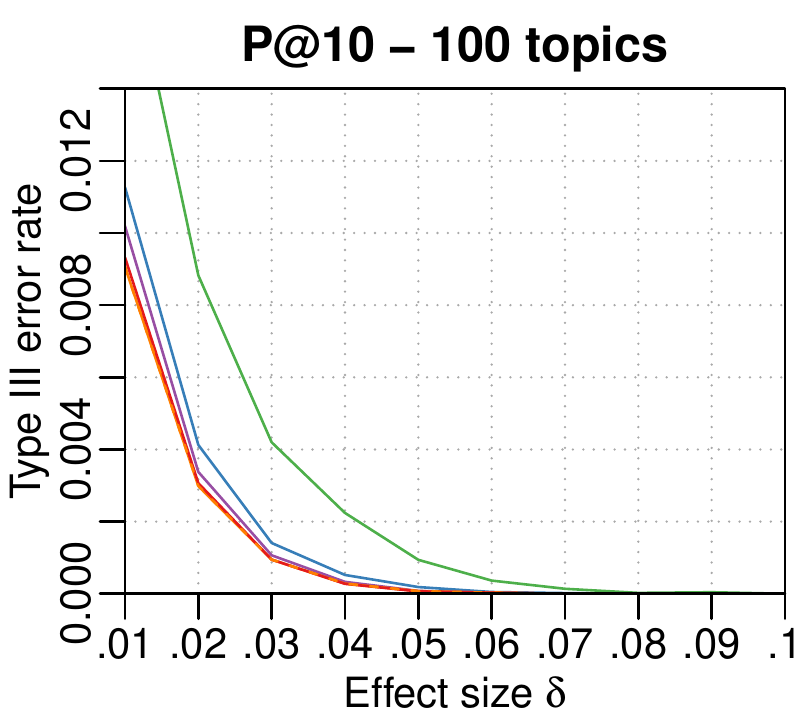}~\includegraphics[scale=\figscale]{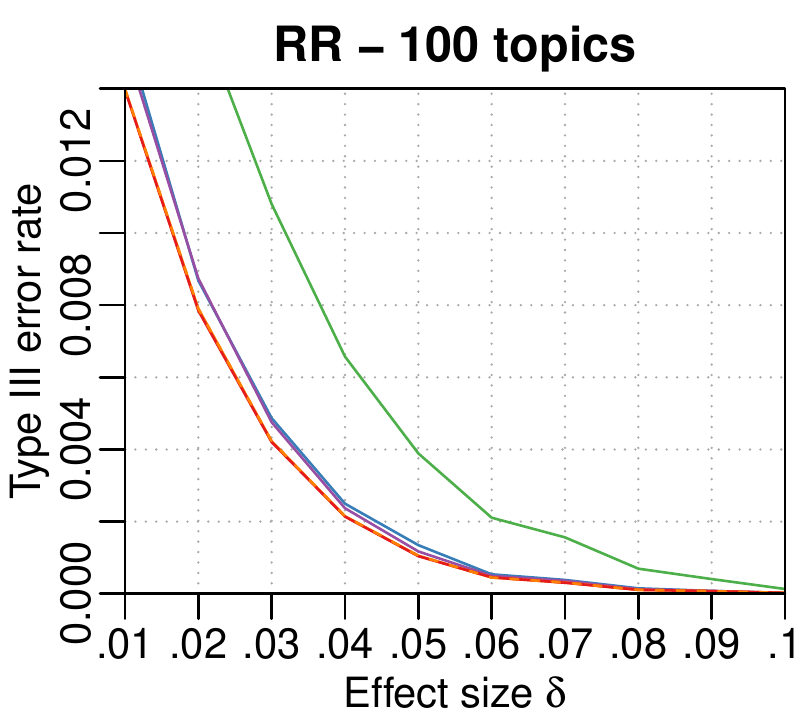}
\caption{Type III error rates in directional 2-tailed tests at $\bm{\alpha=0.05}$. Plots on the same column correspond to the same topic set size, and plots on the same row correspond to the same effectiveness measure. When indiscernible, the $\bm{t}$, permutation and bootstrap tests overlap.}
\label{fig:type3-2tail}
\end{figure*}

\subsection{Type III Errors}\label{sec:results:type3}

The majority of literature on statistical testing is devoted to analyzing the Type I error rate, while the study of Type II errors and power has received considerably less attention. There is however a third type of errors that is virtually neglected in the literature. Type III errors refer to cases in which a wrong directional decision is made after correctly rejecting a non-directional hypothesis\footnote{The related concept of \emph{major conflict} was studied for instance by \citet{voorhees2009redux} and \citet{urbano2013comparison}, but without control over the null hypothesis.}~\citep{kaiser1960directional}. Imagine we observe a positive improvement of our experimental system \sysE over the baseline \sysB, that is, $\overline{D}>0$. If we follow the common procedure of running a 2-tailed test for the non-directional hypothesis $H_0:\mu_B=\mu_E$ and reject it, we typically make the directional decision that $\mu_E>\mu_B$ on the grounds of $\overline{D}>0$ and $p_2\leq \alpha$. In such case, we make a directional decision based on a non-directional hypothesis, but that decision could very well be wrong. In our case, it could be that the baseline system really is better and the set of topics was just unfortunate. That is, $H_0$ would still be correctly rejected, but for the wrong reason. In modern terminology, Type III errors are sometimes referred to as Type S errors (sign)~\citep{gelman2000typeS}.

Using the same data from section~\ref{sec:results:type2}, we can study the rate of Type III errors, coined $\gamma$ by~\citet{kaiser1960directional}. Because our simulated data always had a positive effect $\delta$, we actually count cases in which $\overline{D}<0$ and the non-directional hypothesis was rejected by $p_2\leq\alpha$. In these cases we would incorrectly conclude that the baseline is significantly better than the experimental system. Again, we only report here results at $\alpha=0.05$.

Figure~\ref{fig:type3-2tail} shows the rate of Type III errors as a function of the effect size $\delta$. As anticipated by the Type I error rates, the $t$ and permutation tests lead to much fewer errors than the other tests, with the Wilcoxon and sign tests being again more error-prone and the bootstrap test somewhere in between. The effect of the topic set size is clearly visible again in that the Wilcoxon and sign test become even more unreliable while the others significantly reduce the error rate. In particular, the bootstrap test catches up with the $t$ and permutation tests. Differences across measures are once again caused by the variability of their distributions. $RR$ is by far the most variable measure and hence more likely to observe outliers that lead to wrong conclusions. On the other hand, $ERR@20$ is the most stable measure and the tests show better behavior with it.

We note that the error rates reported in Figure~\ref{fig:type3-2tail} are over the total number of cases. An arguably more informative rate can be computed over the number of significant cases, that is, the fraction of statistically significant results that could actually lead to Type III errors. Even though space constraints do not permit to report these plots, Figures~\ref{fig:power-2tail} and~\ref{fig:type3-2tail} together provide a rough idea.
For instance, with $AP$ and 50 topics, the Type III error rate at the typical $\delta=0.01$ is 0.0069 with the $t$-test. With this effect size, 9.47\% of comparisons come up statistically significant, which means that 7.3\% of significant results could lead to a Type III error.
As another example, with $P@10$ and 50 topics the Type III error rate is 0.0064, with power at 8.9\%. This means that 7.2\% of significant results could lead to erroneous conclusions.

\section{Conclusions}\label{sec:conclusions}

To the best of our knowledge, this paper presents the first empirical study of actual Type I and Type II error rates of typical paired statistical tests with IR data. Using a method for stochastic simulation of evaluation scores, we compared the $t$-test, Wilcoxon, sign, bootstrap-shift and permutation tests on more than 500 million $p$-values, making it also a 10-fold increase over the largest study to date. Our analysis also comprises different effectiveness measures ($AP, nDCG@20, ERR@20, P@10$ and $RR$), topic set sizes (25, 50 and 100), significance levels (0.001 to 0.1), and both 2-tailed and 1-tailed tests, making it the most comprehensive empirical study as well.

Our results confirm that the sign and Wilcoxon tests are unreliable for hypotheses about mean effectiveness, specially with large sample sizes. One could argue that this is because they do not test hypotheses about means, but about medians. However, because of the symmetricity assumption, they are legitimately used as alternatives to the $t$-test with less strict assumptions.
As suggested by previous research, the $t$-test and permutation test largely agree with each other, but the $t$-test appears to be slightly more robust to variations of sample size. On the other hand, the bootstrap-shift test is shown to have a clear bias towards small $p$-values. While this leads to higher power, our results confirm that it also has higher Type I error rates, so there is virtually no gain over the $t$ or permutation tests. This bias decreases with sample size, so the only situation favorable to the bootstrap appears to be that of (very) large test collections. The Wilcoxon test is found in a similar situation of high power but high Type I errors, specially as sample size increases.

An advantage of our study is that it allows us to move past the typical discordance ratios used so far in the literature to compare tests. As discussed here and in previous work, without knowledge of the null hypothesis these discordances do not imply Type I errors. Thanks to our methodology based on simulation, we computed actual Type I error rates and found that both the $t$-test and the permutation test maintain errors at the nominal $\alpha$ level remarkably well, and they do so across measures and sample sizes. The tests are \emph{not} being too conservative as previously suggested, and even though some effectiveness measures are indeed more unstable than others, that does not imply a higher error rate in the tests.

Based on our findings, we strongly recommend the use of the $t$-test for typical hypotheses pertaining to mean system effectiveness, and the permutation test for others. We provide further evidence that the bootstrap-shift test, although with nice theoretical properties, does not behave well unless the sample size is large, so we also recommend its discontinuation.

\begin{acks}
Work carried out on the Dutch national e-infrastructure (SURF Cooperative) and funded by European Union's H2020 programme (770376-2--TROMPA).\\Whichever tree, plant, or bird you're now part of, thank you, Mom.
\end{acks}

%
\bibliographystyle{ACM-Reference-Format}
\balance
\bibliography{2019}

\end{document}